\documentclass[conference]{IEEEtran}
\IEEEoverridecommandlockouts
\usepackage{cite}
\usepackage{amsmath,amssymb,amsfonts}
\usepackage{algorithmic}
\usepackage{graphicx}
\usepackage{subfigure}
\usepackage{booktabs}
\usepackage{comment}
\usepackage{enumerate}
\newtheorem{definition}{Definition}[section]
\usepackage{tabularx}
\usepackage{textcomp}
\usepackage{xcolor}
\usepackage{amsmath}
\usepackage[linesnumbered,commentsnumbered,ruled,vlined]{algorithm2e}
\def\BibTeX{{\rm B\kern-.05em{\sc i\kern-.025em b}\kern-.08em
    T\kern-.1667em\lower.7ex\hbox{E}\kern-.125emX}}
\usepackage{bm}    
\usepackage{tikz} 
\usepackage{ulem}
\usepackage{color}
\newcommand{\PW}[1]{\textcolor{red}{(PW: #1)}}

\begin{document}

\title{Defending Water Treatment Networks:  Cyber Attack Detection via Multi-scale Multi-channel Graph Anomaly Structure}

\title{Defending Water Treatment Networks:  Exploiting Spatio-temporal Effects for Cyber Attack Detection}

\author{\IEEEauthorblockN{1\textsuperscript{st} Dongjie Wang}
\IEEEauthorblockA{
\textit{Department of Computer Science} \\
\textit{University of Central Florida}\\
Orlando,United States \\
wangdongjie@knights.ucf.edu}
\and
\IEEEauthorblockN{2\textsuperscript{nd} Pengyang Wang}
\IEEEauthorblockA{
\textit{Department of Computer Science} \\
\textit{University of Central Florida}\\
Orlando, United States \\
pengyang.wang@knights.ucf.edu}
\and
\IEEEauthorblockN{3\textsuperscript{rd} Jingbo Zhou}
\IEEEauthorblockA{
\textit{Baidu Research} \\
\textit{Baidu Inc.}\\
Beijing, China \\
zhoujingbo@baidu.com}
\and
\IEEEauthorblockN{4\textsuperscript{th} Leilei Sun}
\IEEEauthorblockA{
\textit{Department of Computer Science} \\
\textit{Beihang University}\\
Beijing, China \\
leileisun@buaa.edu.cn}
\and
\IEEEauthorblockN{5\textsuperscript{th} Bowen Du}
\IEEEauthorblockA{
\textit{Department of Computer Science} \\
\textit{Beihang University}\\
Beijing, China \\
dubowen@gmail.com}
\and
\IEEEauthorblockN{6\textsuperscript{th} Yanjie Fu$^{{\dagger}}$}
\IEEEauthorblockA{
\textit{Department of Computer Science} \\
\textit{University of Central Florida}\\
Orlando, United States \\
yanjie.fu@ucf.edu}
}

\maketitle

\begin{abstract}
While Water Treatment Networks (WTNs) are critical infrastructures for local communities and public health, WTNs are vulnerable to cyber attacks. Effective detection of attacks can defend WTNs against discharging contaminated water, denying access,  destroying equipment, and causing public fear.  
While there are extensive studies in WTNs attack detection, they only exploit the data characteristics partially to detect cyber attacks.
After preliminary exploring the sensing data of WTNs, we find that integrating spatio-temporal knowledge, representation learning, and detection algorithms can improve attack detection accuracy.
To this end, we propose a structured anomaly detection framework to defend WTNs by modeling the spatio-temporal characteristics of cyber attacks in WTNs.
In particular, we propose a spatio-temporal representation framework specially tailored to cyber attacks after separating the sensing data of WTNs into a sequence of time segments.
This framework has two key components. 
The first component is a temporal embedding module to preserve temporal patterns within a time segment by projecting the time segment of a sensor into a temporal embedding vector. 
We then construct Spatio-Temporal Graphs (STGs), where a node is a sensor and an
attribute is the temporal embedding vector of the sensor, to describe the state of the WTNs.
The second component is a spatial embedding module, which learns the final fused embedding of the WTNs from STGs. 
In addition, we devise an improved one class-SVM model that utilizes a new designed pairwise kernel to detect cyber attacks. 
The devised pairwise kernel augments the distance between normal and attack patterns in the fused embedding space. 
Finally, we conducted extensive experimental evaluations with real-world data to demonstrate the effectivness of our framework: it achieves an accuracy of $91.65\%$, with average improvement ratios of $82.78\%$  and $22.96\%$ with respect to F1 and AUC,  compared with  baseline methods.

\end{abstract}

\section{Introduction}

\begin{figure}[htbp]
    \centering
    \includegraphics[width=\linewidth]{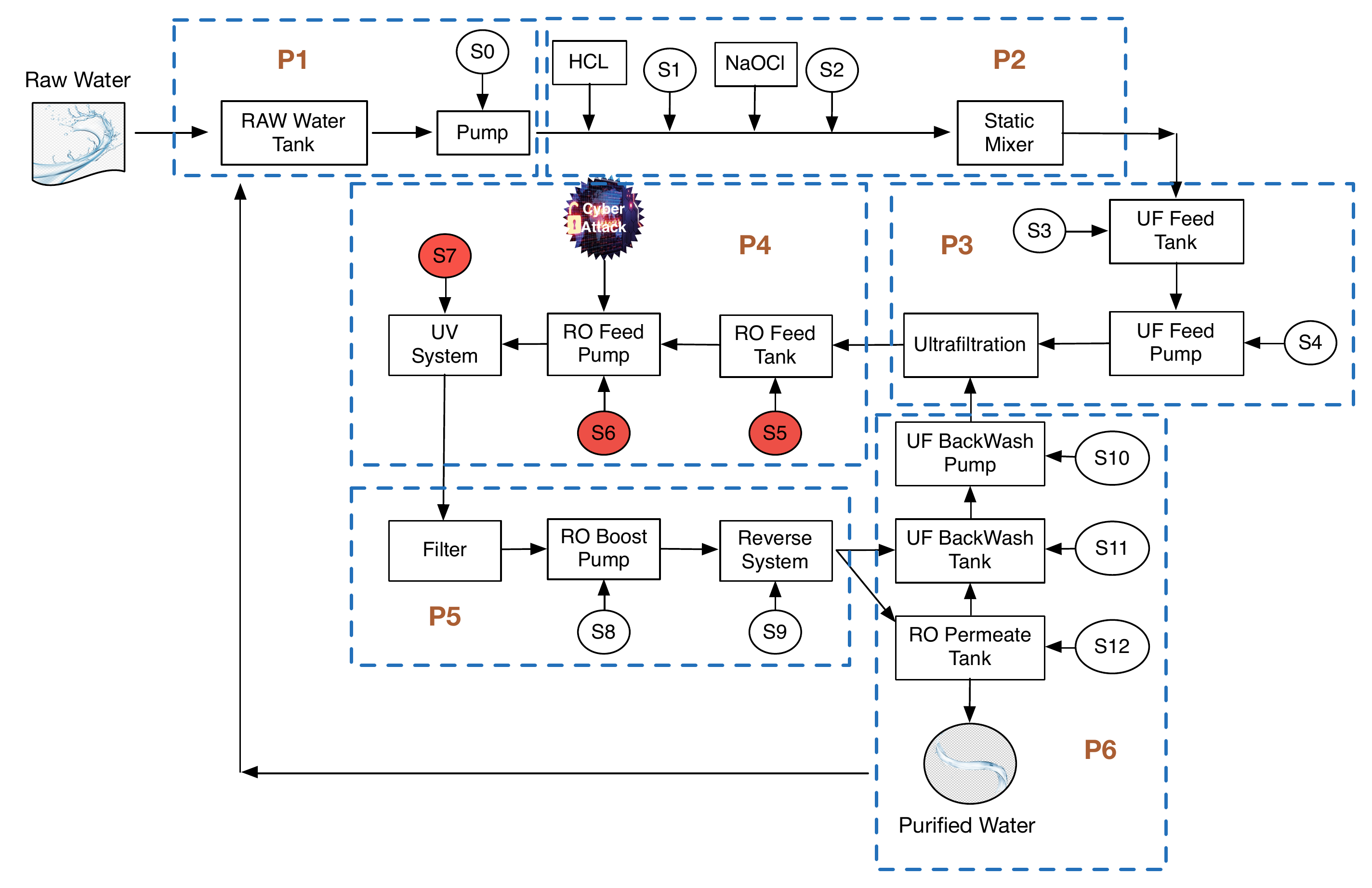}
    \caption{ 
    Cyber attack example: one cyber attack happens at RO Feed Pump of P4, then the cyber attack effect spreads to other devices in P4.}
     \vspace{-0.4cm}
    \label{fig:wtn}
\end{figure}

Water Treatment Networks (WTNs) are critical infrastructures that utilize industrial control systems, sensors and communication technologies to control the water purification processes to improve the water quality and distribution for drinking, irrigation, or industrial uses. Although it is a critical infrastructure,
WTNs are vulnerable to cyber attacks. 
For example, the water sector reported the fourth largest number of incidents in 2016~\footnote{https://www.osti.gov/servlets/purl/1372266}. 
How does a cyber attack to WTNs look like?
Figure~\ref{fig:wtn} shows that a water treatment procedure includes six stages ({\it i.e.}, P1-P6), each of which is monitored by sensors; a cyber attack compromises the RO Feed Pump sensor of P4 to change the levels of chemicals being used to treat tap water. 
As a result, there is a compelling need for an effective solution to attack detection in WTNs.

In the literature, there are a number of studies about cyber attack detection in WTNs~\cite{goh2017anomaly,romano2010real,li2019mad,feng2017multi}.
However, most of these studies only exploit traditional spaiotemporal data preprocessing and pattern extraction methods to distinguish attack patterns.
Our preliminary explorations find that tremendous opportunities exist in solving the problem by teaching a machine to augment the differences between normal and attack patterns.
To this end, in this paper, we aim to effectively solve the attack detection problem by augmenting the difference between normal and attack patterns in WTNs. 

However, it is challenging to mine the spatio-temporal graph stream data of WTNs and identify the strategy to maximize the pattern divergence between normal  and attack behaviors. 
By carefully examining the sensing data of WTNs, we identify three types of characteristics of cyber attacks: 
(1) {\it delayed effect}, meaning that many attacks will not take effects immediately, but usually exhibit symptoms after a while;
(2) {\it continued effect}, meaning that the effects of attacks will sustain for a while, not disappear rapidly;
(3) {\it cascading effect}, meaning that the effects of attacks  propagate to other sensors across the whole WTNs. 
Specifically, the {\it delayed } and {\it continued} effects are both temporal, and the {\it cascading effect} is spatial. 
More importantly, these three effects are mutually coupled, impacted, and boosted in WTNs.
A new framework is required to address the margin maximization between normal and attack pattern learning, under the three coupled effects. 

Along this line, we propose a structured detection framework.
This framework has two main phases:
(1) spatio-temporal representation learning, which includes two modules: incorporating temporal effects and  spatial effects;
(2) improved unsupervised one-class detection, which utilizes a new designed pairwise kernel to make detection more accurate. 
Next, we briefly introduce our structured spatio-temporal detection framework named STDO.

{\bf Phase 1: Spatio-temporal representation learning.}
This phase aims to learn a good spatio-temporal representation over the sensing data of WTNs with two modules.
The first module of this part is to integrate temporal effects.
Cyber attacks in WTNs exhibit temporally-dependent attack behaviors, sequentially-varying attack purposes over time, and delayed negative impacts. 
Traditional methods ( e.g., AR, MA, ARMA, ARIMA, arrival density of point process, change point detection) mostly model the patterns of data points at each timestamp.
However, we identify that partitioning the sensing data into a sequence of time segments can help to better describe {\it delayed} and {\it continued effect} of attacks. 
Therefore, we propose to segment the sensing data into a sequence of time segments. 
We then exploit a sequence-to-sequence (seq2seq) embedding method to characterize the temporal dependencies within each time segment. 
To improve the seq2seq method, we develop a new neural reconstruction structure to reconstruct not just a time segment, but also first and second derivatives of momentum of  the time segment. 
In this way, the improved seq2seq method can have the awareness of  values, acceleration, and jerk (second order derivatives) of sensor measurements. 
Through this module, we obtain the temporal embedding of each time segment of each  sensor.

The second module is to integrate spatial effects.
The effects of cyber attacks in WTNs will spatially diffuse and propagate to other sensors over time.
Therefore, exploring the  propagation structure  can significantly model attack patterns and improve detection accuracy.
However, how can we capture the spatial structures of propagation?
The topology of WTNs is a graph of interconnected sensors.
We map the temporal embedding of one time segment of a sensor to the graph of WTNs as node attributes.
We construct the Spatio-Temporal Graphs (STGs), where a node is a sensor and an attribute is the temporal embedding of the sensor, to describe the state of the WTNs.
In this way, the STGs not only contain spatial connectivity among different sensors, but also include temporal patterns by mapping temporal embedding. 
We develop graph embedding model to jointly learn the state representations of the WTNs from the relational STGs.  

{\bf Phase 2: Improving Unsupervised One-Class Detection Model.} 
In reality, WTNs are mostly normal yet with a small number of cyber attack events, so the attack data samples are quite rare. This trait makes the data distributions of normal and attack extremely imbalanced. How can we overcome the imbalanced data distribution to accurately detect cyber attack? One-class detection fits well this problem. In particular, one-class SVM (OC-SVM) is a popular detection model, in which a hyper-plane is identified to divide normal and abnormal data samples after being mapped to a high-dimmensional space by kernel functions. While vanilla OC-SVM achieves promising results, the kernel functions can be improved by exploiting the similarities between data samples.
Specifically, we propose a new pair-wise kernel function to augment the distance between normal and attack patterns by preserving similarity across different data samples. Consequently, normal data samples are grouped into a cluster, while abnormal data samples are pushed away from normal data.
In our framework, we feed the learned state representations of the WTN into the improved OC-SVM to use the pairwise kernel to detect attacks.

In summary, we develop a structured detection framework for cyber attack detection in WTNs. 
Our contributions are as follows:
(1) we investigate an important problem of defending critical graph-structured infrastructures via cyber attack detection, which is important for building resilient and safe communities. 
(2) we develop a structured detection framework to maximize the margin between normal and attack patterns, by integrating spatio-temporal knowledge, deep representation learning, and pairwise one-class detection. 
(3) we implement and test our proposed framework with real-world water treatment network data and demonstrate the enhanced performance of our method.
Specifically, our detection method achieves an accuracy of $91.65\%$, with average improvement ratios of $82.78\%$  and $22.96\%$ with respect to F1 and AUC,  compared with baseline methods.

\section{Problem Statement and Framework Overview}
We first introduce the statement of the problem, and then present an overview of the framework.

\begin{figure*}[htbp]
\setlength{\abovecaptionskip}{-0.01cm} 
\centering
\subfigure[\textbf{P1:} Embedding time segments sequential patterns.]{
\begin{minipage}[t]{0.33\linewidth}
\centering
\includegraphics[width=1\linewidth]{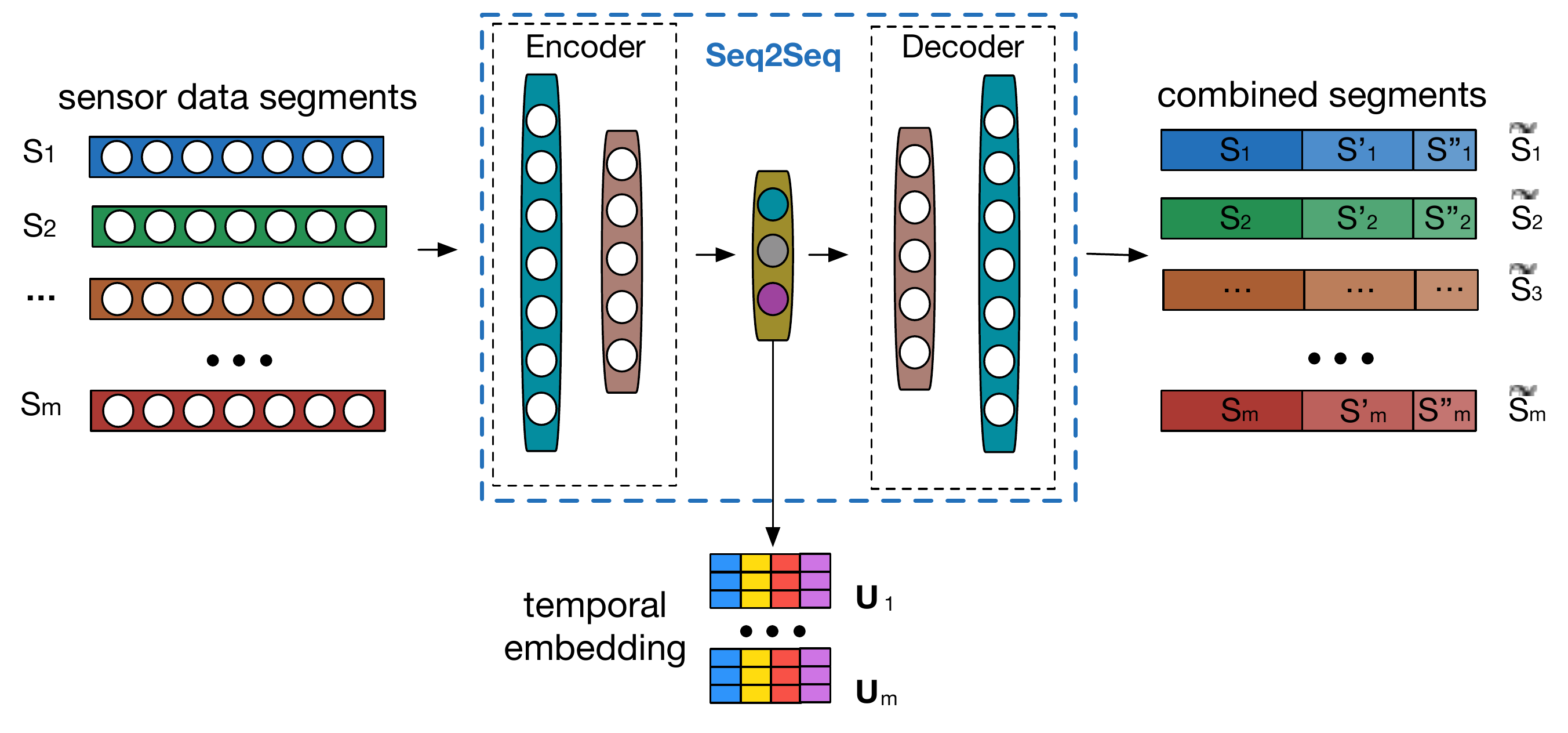}
\label{fig:temporal_embedding}
\end{minipage}%
}%
\subfigure[\textbf{P1:} Modeling spatio-temporal patterns over STG.]{
\hspace{0.5cm}
\begin{minipage}[t]{0.33\linewidth}
\centering
\includegraphics[width=1\linewidth]{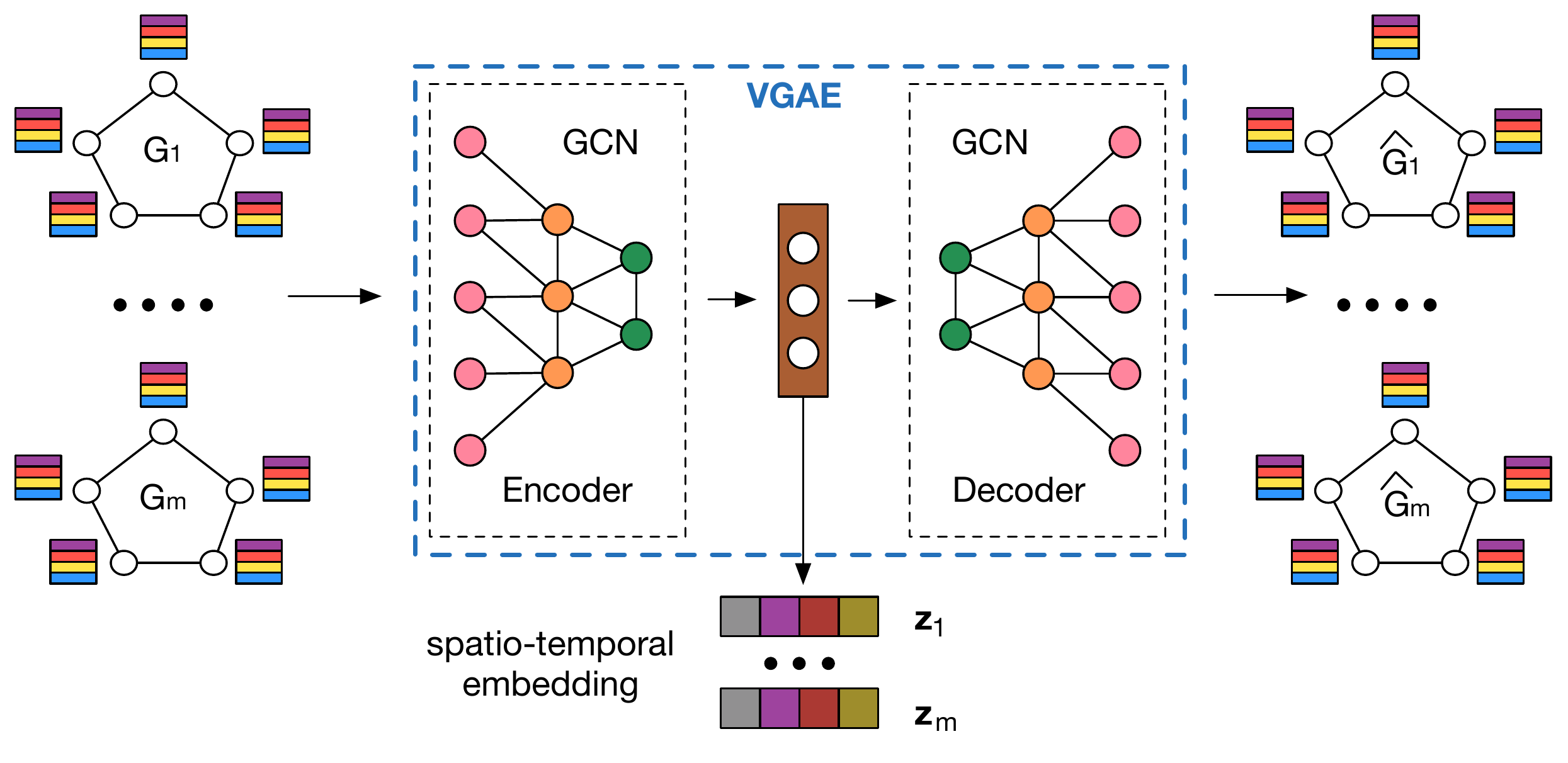}
\label{fig:collective_embedding}
\end{minipage}%
}%
\subfigure[\textbf{P2:} Anomaly detection with data similarity.]{
\hspace{-0.6cm}
\begin{minipage}[t]{0.33\linewidth}
\centering
\includegraphics[width=0.75\linewidth]{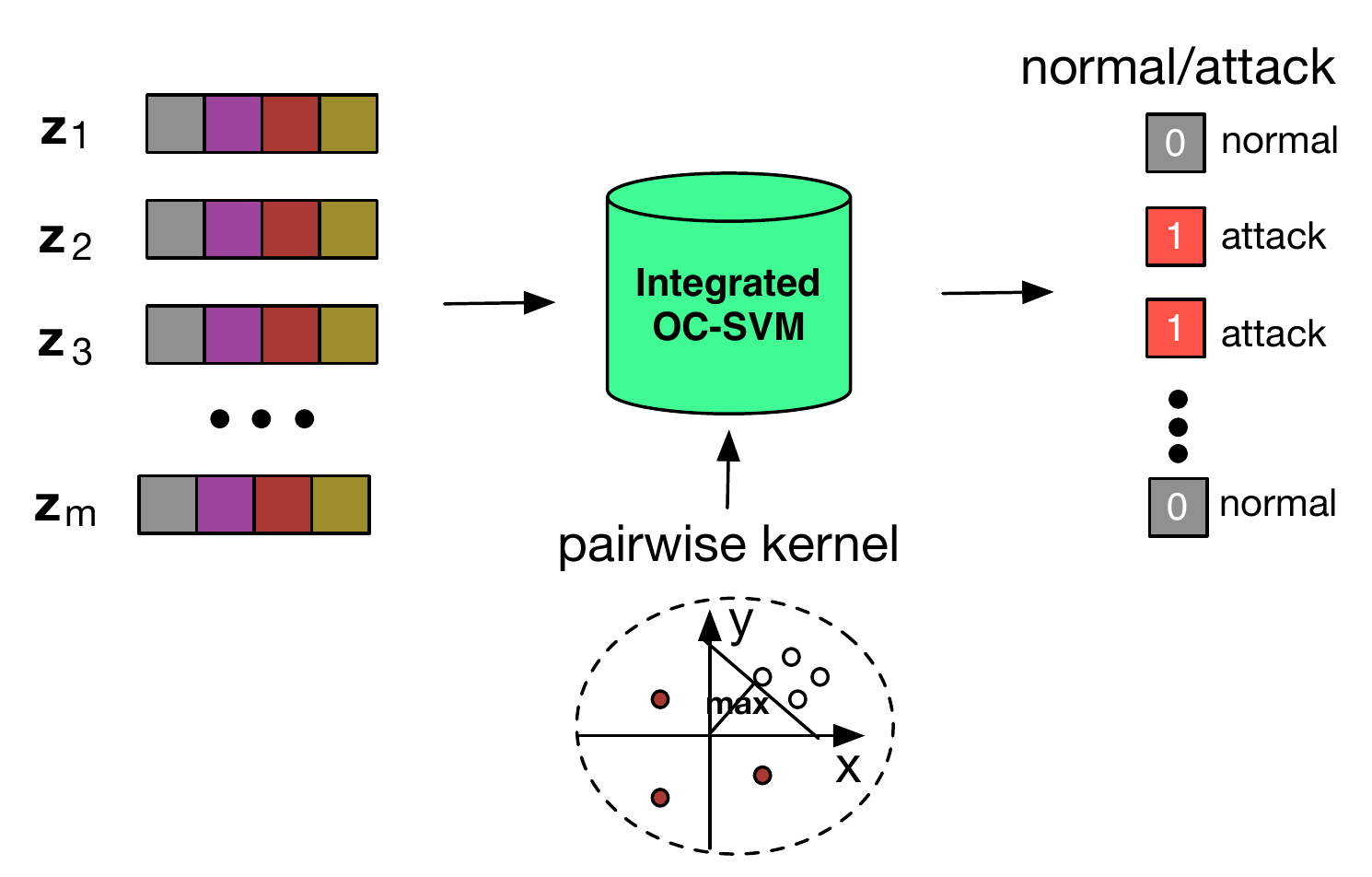}
\vspace{0.39cm}
\label{fig:outlier_detection}
\end{minipage}
}%

\centering
\caption{The overview of cyber attack detection framework in water treatment network}
\label{fig:framework}
\vspace{-0.6cm}
\end{figure*}

\subsection{Problem Statement}
We aim to study the problem of cyber attack detection using the sensing data of WTNs. 
We observe that cyberattacks in WTNs exhibit not just spatial diffusion patterns, but also two temporal effects (i.e., delayed and continued).
As a result, we partition the sensing data streams into non-overlapped time segments. 
We investigate detection on the time segment level.

\begin{definition}	
{\it The WTN Attack Detection Problem. }
Formally, assuming a WTN consists of $N$ sensors, given the sensing data streams of a WTN, we evenly divide the streams into $m$ non-overlapped segments by every $K$ sensory records.
Consequently, we obtain a segment sequence  $\mathbf{S}=[\mathbf{S}_1, \mathbf{S}_2, \cdots, \mathbf{S}_i, \cdots, \mathbf{S}_m]$, where the matrix $\mathbf{S}_i \in \mathbb{R}^{N \times K}$ is the $i$-th segment.
Each segment is associated with a cyber attack status label:  if a cyber attack happens within $\mathbf{S}_i$, the label of this segment is marked as $y_i = 1$; Otherwise, $y_i = 0$.
The objective is to develop a framework that takes the segment sequence $\mathbf{S}$ as inputs,  and output the corresponding cyber attack labels for each segment to maximize detection accuracy.
\end{definition}

\subsection{Framework Overview}
Figure \ref{fig:framework} shows that our framework includes two phases:
(1) Spatio-temporal representation learning (\textbf{P1}); 
(2) Improving unsupervised one-class detection (\textbf{P2}).
Specifically, there are two modules in Phase 1:
(a) Embedding time segment sequential patterns, in which a seq2seq model is leveraged to capture the temporal dependencies within a segment.
Later, the learned representations are attached to each node (sensor) in the WTNs as node attributes to construct STGs. Be sure to notice that temporal patterns are integrated by attaching temporal embeddings as attributes; and spatial connectivity is integrated via a graph structure of sensors, which is introduced next. 
(b) Modeling spatio-temporal patterns over STGs, in which the fused embedding is learned through an encode-decode paradigm integrated with Graph Convolutional Network (GCN). 
The fused embedding summarizes the information of STGs to profile the spatio-temporal characteristics in WTNs.
Finally, the Phase 2 exploit the fused embedding as inputs to detect attacks.
The Phase 2 has one module, namely anomaly detection with pairwise segment similarity awareness. 
Specifically,  the fused embedding is fed into an improved one-class anomaly detector integrated with awareness of pairwise segment similarity.
Specifically, the similarities between two different segment embedding vectors are introduced to the pairwise kernel of the detector to augment the distance between normal and attack patterns.

\section{Proposed Method}
We first introduce time segment embedding, then illustrate the construction of spatio-temporal graphs using temporal embedding and sensor networks, present spatio-temporal graph-based representation learning, and, finally, discuss the integration with one-class detection.

\subsection{Embedding Time Segments Sequential Patterns}
\label{sec:temporal_embedding}
We first model the sequential patterns of time segments.
The sequential patterns of WTN involves two essential measurements: (1) changing rate and (2) the trend of changing rate, which correspond to the first and second derivatives respectively.
Therefore, in addition to the raw data points in one segment, we introduce both the first and second derivatives to quantify the sequential patterns, resulting in an augmented segment. Formally, blow we define the augmented segment.
\begin{definition}
{\it Augmented Segment.} Given a sensory data segment denoted by  $\mathbf{S}_i=[\mathbf{v}_i^1, \mathbf{v}_i^2, \cdots, \mathbf{v}_i^k, \cdots, \mathbf{v}_i^K]$, where $\mathbf{v}_i^k \in \mathbb{R}^{N\times 1}$ denotes the sensory measurements of all the sensors of the $i$-th segment at the $k$-th record.
Then, the corresponding first-order derivative segment $\mathbf{S}_i^{'}$ is $\mathbf{S}_i^{'}=[\frac{\partial \mathbf{S}_i}{\partial \mathbf{v}_i^2}, \frac{\partial \mathbf{S}_i}{\partial \mathbf{v}_i^3}, \cdots, \frac{\partial \mathbf{S}_i}{\partial \mathbf{v}_i^K}]$, and the corresponding second-order derivative segment $\mathbf{S}_i^{''}$ is $\mathbf{S}_i^{''}=[\frac{\partial \mathbf{S}_i^{'}}{\partial \mathbf{v}_i^3}, \frac{\partial \mathbf{S}_i^{'}}{\partial \mathbf{v}_i^4}, \cdots, \frac{\partial \mathbf{S}^{'}}{\partial \mathbf{v}_i^K}]$.
The augmented segment $\tilde{\mathbf{S}}_i$ is then defined as the concatenation of the raw segment, the first-order  derivative segments, and the second-order  derivative segments: $\tilde{\mathbf{S}}_i = [\mathbf{S}_i, \mathbf{S}_i^{'}, \mathbf{S}_i^{''}]$.
For convenience, $\tilde{\mathbf{S}}_i$ also can be denoted as $\tilde{\mathbf{S}}_i=[\mathbf{r}_i^1, \mathbf{r}_i^2,\ldots,\mathbf{r}_i^{3K-3}]$, where 
elements in $[\mathbf{r}_i^1, \mathbf{r}_i^2, \cdots, \mathbf{r}_i^{K}]$ corresponds to each element in $\mathbf{S}_i$, 
elements in $[\mathbf{r}_i^{K+1}, \mathbf{r}_i^{K+2}, \cdots, \mathbf{r}_i^{2K-1}]$ corresponds to each element in $\mathbf{S}^{'}_i$, 
and the elements in $[\mathbf{r}_i^{2K}, \mathbf{r}_i^{2K+1}, \cdots, \mathbf{r}_i^{3K-3}]$ corresponds to each element in $\mathbf{S}^{''}_i$ respectively.
\end{definition}

We here provide an example of constructing an augmented segment. 
Suppose there are two sensors in WTNs, there are three measurement records in each time segment.
In such WTN, $N$ is 2 and $K$ is 3.
Considering the $i$-th segment $\mathbf{S}_i = [[1,3,4],[2,8,5]]$, the size of $\mathbf{S}_i$ is $2\times3$. 
We then calculate the $\mathbf{S}^{'}_i$ by row: $\mathbf{S}^{'}_i = [[2,1],[6,-3]]$.
Afterward, $\mathbf{S}^{''}_i=[[-1],[-9]]$.
Finally, we concatenate these three segments by row: $\tilde{\mathbf{S}}_i =
[[1,3,4,2,1,-1]
[2,8,5,6,-3,-9]]
$.

Figure \ref{fig:temporal_embedding} shows the process of temporal embedding.
The temporal embedding process develops a seq2seq model based on the encoder-decoder paradigm that takes a non-augmented segment as inputs, and reconstructs the corresponding augmented segment.
The objective is to minimize the loss between the original augmented segment and the reconstructed one.
Next, we provide an example about how our model operate the $i$-th segment $\mathbf{S}_i$.

The encoding step feeds the segment $\mathbf{S}_i$ into a seq2seq encoder, and outputs the latent representation of the segment $\mathbf{U}_i$.
Formally,  as shown in Equation~\ref{equ:seq_enc}, given the segment data $\mathbf{S}_i=[\mathbf{v}_i^1, \mathbf{v}_i^2, \ldots, \mathbf{v}_i^K]$,
the first hidden state $\mathbf{h}^1$ is calculated by the first time step value. 
Then recursively, the hidden state  of the previous time step $\mathbf{h}^{t-1}$ and the current time step value $\mathbf{v}_i^t$ are fed into a LSTM model to produce the current time step hidden state $\mathbf{h}^t$.
Finally, we concatenate all of the hidden states by row (sensor) to obtain the latent feature matrix $\mathbf{U}_{i}$.
\begin{equation}
\left\{
             \begin{array}{lr}
             \mathbf{h}^1 =\sigma (\mathbf{W}_e \mathbf{v}_i^1 + \mathbf{b}_e), &  \\
             \mathbf{h}^t=LSTM([\mathbf{v}_i^t, \mathbf{h}^{t-1}]), \forall t \in \{2,\dots,K\}, \\
             \mathbf{U}_{i} = CONCAT(\mathbf{h}^1, \mathbf{h}^2, \ldots , \mathbf{h}^K), &  
             \end{array}
\right.
\label{equ:seq_enc}
\end{equation}
where $\mathbf{W}_e$ and $\mathbf{b}_e$ are the weight and bias of the encoding step, respectively.

In the decoding step, the decoder takes $\mathbf{U}_i$ as inputs and generates a reconstructed augmented segment: $[\mathbf{\hat{r}}_i^1,\mathbf{\hat{r}}_i^2, \ldots, \mathbf{\hat{r}}_i^{3K-3}]$. 
Formally, as shown in Equation~\ref{equ:decoder}, the first hidden state $\mathbf{\hat{h}}^1$ of the decoder is copied from the last hidden state of encoder $\mathbf{h}^K$. 
Then, the previous time step hidden state $\mathbf{\hat{h}}^{t-1}$, the previous time step element $\mathbf{\hat{r}}_i^{t-1}$, and the latent feature vector $\mathbf{U}_i$ are input into the LSTM model to produce the current time step hidden state $\mathbf{\hat{h}}^t$.
Finally, the reconstructed value of current time step $\mathbf{\hat{r}}_i^{t}$  is produced by current hidden state $\mathbf{\hat{h}}^t$ that is activated by sigmoid function $\sigma$. 
\begin{equation}
\left\{
             \begin{array}{lr}
             \hat{\mathbf{h}}^1 = \mathbf{h}^K, \\
            \hat{\mathbf{r}}_i^{1} = \sigma(\mathbf{W}_d \hat{\mathbf{h}^{1}} + \mathbf{b}_d), \\
              \hat{\mathbf{h}}^{t} = LSTM([
             \hat{\mathbf{r}}^{t-1}_i, \hat{\mathbf{h}}^{t-1},
             \mathbf{U}_i
             ]), \forall t \in \{2,\dots,K\},  \\
            \hat{\mathbf{r}}_i^{t} = \sigma(\mathbf{W}_d \hat{\mathbf{h}^{t}} + \mathbf{b}_d),  \forall t \in \{2,\dots,K\}, \\

             \end{array}
\right.
\label{equ:decoder}
\end{equation}
where $\mathbf{W}_d$ and $\mathbf{b}_d$ are the weight and bias for the decoding step respectively.

After the decoding step, we obtain the reconstructed augmented segment sequence $[\hat{\mathbf{r}}_i^1, \hat{\mathbf{r}}_i^2, \ldots, \hat{\mathbf{r}}_i^{3K-3}]$. 
The objective is to minimize the reconstruction loss between the original and reconstructed augmented segment sequence.
The overall loss is denoted as 
\begin{equation}
    \min 
    \sum \limits_{i=1}^{m} \sum \limits^{3K-3}_{k=1} || \mathbf{r}_i^k - \hat{\mathbf{r}}_i^{k} ||^2.
\end{equation}

Along this line, we obtain the latent temporal embedding at the $i$-th time segment, denoted by $\mathbf{U}_i$.

\subsection{Temporal Embedding as Node Attributes: Constructing Spatio-temporal Graphs}

\begin{figure}[htbp]
    \centering
    \includegraphics[width=\linewidth]{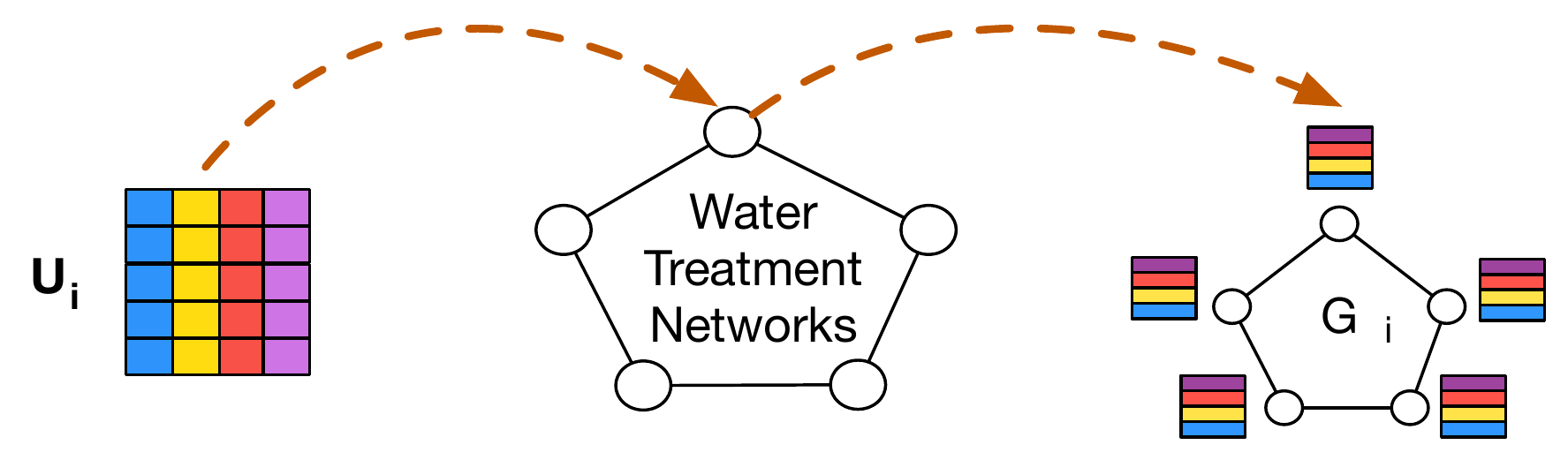}
    \caption{The illustration of constructing spatio-temporal graphs.}
    \label{fig:graph_map}
     \vspace{-0.3cm}
\end{figure}

The temporal embedding, obtained by Section \ref{sec:temporal_embedding}, describes and models the temporal effects of cyber attacks.
Then, to further incorporate the spatial effects of WTNs, we map the temporal embedding to WTNs as node attributes.
Taking the temporal embedding of the $i$-th segment $\mathbf{U}_i$ as an example.
Since each row of  $\mathbf{U}_i$ is a temporal embedding of a segment of one senor (node), we mapped each row of $\mathbf{U}_i$ to the corresponding node (sensor) as attributes, resulting in an attributed WTNs $G_i$, which we call a {\it Spatio-temporal Graph} (STG) that preserves both the spatial and temporal effects.


\subsection{Learning Representations of STGs}
Figure \ref{fig:collective_embedding} shows that we develop a spatiotemporal graph representation learning framework to preserve not just temporal patterns, but also spatial patterns in a latent embedding space.
We take the STG of the $i$-th time segment, $G_i$, as an example to explain the representation process.
Formally, we denote $G_i$ by $G_i = (\mathbf{U}_i,\mathbf{A}_i)$, where $\mathbf{A}_i$ is an adjacency matrix that describes the connectivity among different sensors;
$\mathbf{U}_i$ is a feature matrix that is formed by the temporal embedding of all the sensors of the $i$-th time segment.
The representation learning process is formulated as: given the $i$-th STG $G_i$, the objective is to minimize the reconstruction loss between the input $G_i$ and the reconstructed graph $\hat{G}_i$, by an encoding-decoding framework, in order to learn a latent embedding $\mathbf{z}_i$.

The neural architecture of the encoder includes two Graph Convolutional Network (GCN) layers. 
The first GCN layer take $\mathbf{A}_i$ and $\mathbf{U}_i$  as inputs, and then outputs the lower-dimensional feature matrix $\mathbf{\hat{U}}_i$.
Specifically, the encoding process of the first GCN layer is given by: 
\begin{align}
\begin{split}
    \mathbf{\hat{U}}_i &= 
    RELU(GCN(\mathbf{U}_i,\mathbf{A}_i)) \\
    &=RELU(\hat{\mathbf{D}}_i^{-\frac{1}{2}}
    \mathbf{A}_i\hat{\mathbf{D}}_i^{-\frac{1}{2}}\mathbf{U}_i\mathbf{W}_0)
\end{split}
\end{align}
where $\hat{\mathbf{D}_i}$ is the diagonal degree matrix of $G_i$, and  $\mathbf{W_0}$ is the weight matrix of the first GCN layer.

Since the latent embedding $\mathbf{z}_i$ of the graph is sampled from one prior normal distribution, here the purpose of the second GCN layer is to assess the parameters of the prior distribution.
This layer takes $\mathbf{A}_i$ and $\mathbf{\hat{U}}_i$ as the input, then produces the mean value $\bm{\mu}$ and the variance value $\bm{\delta}^2$ of the prior normal distribution as the output. 
Thus the encoding process of the second GCN layer can be formulated as 
\begin{equation}\label{eqn:secondlayer}
    \bm{\mu},log(\bm{\delta}^2) = GCN(
    \hat{\mathbf{U}}_i,\mathbf{A}_i)
     = \hat{\mathbf{D}}_i^{-\frac{1}{2}}\mathbf{A}_i\hat{\mathbf{D}}_i^{-\frac{1}{2}}
     \hat{\mathbf{U}}_i\mathbf{W}_1,
\end{equation}
where $\mathbf{W_1}$ is the weight matrix of the second GCN layer.
Then we utilize the reparameterization trick to mimic the sample operation to construct the latent representation $\mathbf{z}_i$.
The process is formulated as 
\begin{equation}
    \mathbf{z}_i=\bm{\mu}+\bm{\delta} \times \epsilon,
\end{equation}
where $\epsilon \sim \mathcal{N}(0,1)$.

The decoding step takes the latent representation $\mathbf{z}_i$ as the input and outputs the the reconstructed adjacent matrix $\mathbf{\hat{A}}_i$.
The decoding process is denoted as 
\begin{equation}
    \mathbf{\hat{A}}_i = \sigma(\mathbf{z}_i\mathbf{z}_i^T).
\end{equation}
In addition, the core calculation of the decoding step can be denoted as $\mathbf{z}_i\mathbf{z}_i^T = \left \|\mathbf{z}_i\right \| \left \|\mathbf{z}_i^T\right\| \cos\theta$.
Owing to the $\mathbf{z}_i$ is the node level representation, the inner product calculation is helpful to capture the correlation among different sensors.

We minimize the joint loss function $\mathcal{L}_{g}$ during the training phase, which is formulated as Equation \ref{equ:loss}.
$\mathcal{L}_{g}$ includes
two parts. 
The first part is Kullback-Leibler divergence between the distribution of $\mathbf{z}_i$ and the prior standard normal distribution denoted by $\mathcal{N}(0,1)$.
The second part is the squared error between $\mathbf{A}_i$ and $\mathbf{\hat{A}}_i$.
Our training purpose is to make the  $\mathbf{\hat{A}}_i$ as similar as $\mathbf{A}_i$, and to let the distribution of $\mathbf{z}_i$ as close as $\mathcal{N}(0,1)$.
The total loss is denoted as 


\begin{equation}
        \mathcal{L}_{g} = \sum \limits_{i=1}^{m} \underbrace{
        KL[q(\mathbf{z}_i|\mathbf{X}_i,\mathbf{A}_i) || p(\mathbf{z}_i)]
        }_{\text{KL Divergence between $q(.)$ and $p(.)$}}
        +
        \overbrace{
        \sum_{j=1}^{w} \left \| \mathbf{A}_i-\hat{\mathbf{A}}_i \right \|^2 
        }^{\text{Loss between $\mathbf{A}_i$ and $\mathbf{\hat{A}}_i$}}
    \label{equ:loss}
\end{equation}

When the model converges, we apply the global average aggregation to $\mathbf{z}_i$. 
Then the $\mathbf{z}_i$ becomes the graph-level representation of the WTNs, which contains the spatio-temporal information of the whole system at $i$-th time segment.

\subsection {One-Class Detection with Data Similarity Awareness}
In reality, most of sensor data are normal, and attacks related data are scarce and expensive.
This indeed results into the problem of unbalanced training data.
How can we solve the problem?
Can we develop a solution that only uses normal data for attack detection?
This is the key algorithm challenge for this phase. 
One-class classification is a promising solution that aims to find a hyperplane to distinguish normal and attack patterns only using normal data.
Specifically, OC-SVM is a classical one-class classification model. OC-SVM includes two steps:
(1) mapping low dimensional data into a high dimensional feature space by a kernel function. 
(2) learning the parameters of hyper-plane to divide normal and abnormal data via  optimization.

Intuitively, in the hyperspace provided by OC-SVM, the normal (or abnormal) data are expected to be closer, while there should be a large distance between normal and abnormal data.
In other words, similar data points should be closer to each other than dissimilar ones.
However, traditional kernel functions ({\it e.g.}, linear, nonlinear, polynomial, radial basis function (RBF), sigmoid) cannot preserve such characteristic well. 
How can we make  data samples well-separated in order to achieve such characteristic?  
To address the question, we propose a new pairwise kernel function that is capable of reducing the distances between similar data points, while maximizing the distances between dissimilar ones.
Formally, given the representation matrix $\mathbf{Z} = [\mathbf{z}_1, \cdots, \mathbf{z}_i, \cdots, \mathbf{z}_m]$, the pairwise kernel function is given by :
\begin{equation}
    Kernel = \tanh \left(\frac{1}{\mathcal{D}( \mathbf{Z})}\mathbf{Z}\mathbf{Z}^{T}+sim(\mathbf{Z}, \mathbf{Z}^{T})+\mathbf{c}\right)
    \label{con:newk}
\end{equation}
where $\mathbf{Z}^{T}$ is the transpose of $\mathbf{Z}$, $\mathcal{D}(\mathbf{Z})$ is the covariance matrix of $\mathbf{Z}$, and $sim(\mathbf{Z}, \mathbf{Z}^T) \in \mathbb{R}^{N \times N}$ is the pairwise similarity matrix between segments. 
Compared with the vanilla sigmoid kernel function, we add $sim(\mathbf{Z}, \mathbf{Z}^T)$, where the  range of $sim(\mathbf{Z}, \mathbf{Z}^T)$ is $[-1,1]$.
If two segments are more similar, the corresponding value in $sim(\mathbf{Z}, \mathbf{Z}^T)$ is closer to $1$. 
Otherwise, the value is closer to $-1$.
Therefore, when two segments are similar (e.g., both are normal or abnormal samples), the proposed parwise kernel function will push these two segments closer; otherwise, these two segments will be set away from each other.

\begin{figure}[htbp]
    \centering
    \includegraphics[width=0.35\textwidth]{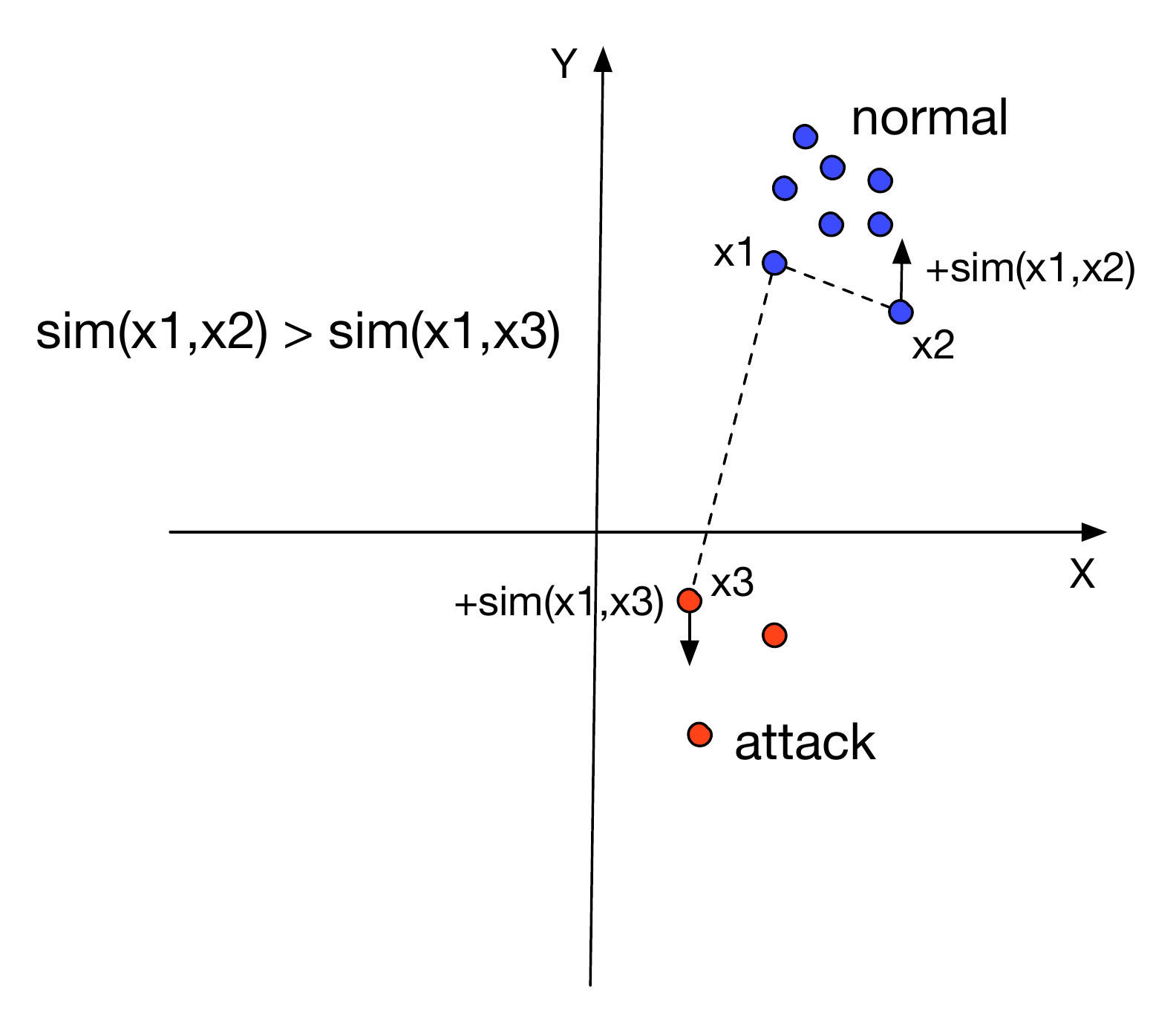}
    \setlength{\abovecaptionskip}{-0.01cm}  
    \caption{The illustration of pairwise kernel, given normal data $x_1$, owing to $x_2$ is normal and $x_3$ is attack, $sim(x_1,x_2) \textgreater sim(x_1,x_3)$ and the directions of $sim(x_1,x_2)$ and $sim(x_1,x_3)$ are opposite. Pairwise kernel increase the distance between $x_2$ and $x_3$ .}
    \label{fig:ker_func}
     \vspace{-0.2cm}
\end{figure}

The pairwise kernel is able to enlarge the distance among different category samples in feature space, which makes the OC-SVM converge more easily and detect cyber attacks more accurate.
Figure \ref{fig:outlier_detection} shows the detection process of  cyber attacks.
The spatio-temporal embedding $\mathbf{z}_i$ is fed into the  integrated OC-SVM to detect cyber attacks by utilizing the pairwise kernel function, and to output the corresponding status labels of WTNs, to indicate whether a cyber attack happen or not at the $i$-th time segment.

\subsection{Comparison with Related Work}
Recently, lots of attempts have been made to detect cyber attacks in WTNs.
For instance, Lin \textit{et al.} utilized a probabilistic graphical model to preserve the spatial dependency among sensors in WTNs and a one-class classifier to detect cyber attacks \cite{lin2018tabor}.
Li \textit{et al.} regarded the LSTM and RNN as the basic model of the GAN framework to develop an anomaly detection algorithm to detect cyber attacks in WTNs \cite{li2019mad}.
Raciti \textit{et al.} constructed one real-time anomaly detection system based on cluster model \cite{raciti2012anomaly}.
However, these models exhibit several limitations when detecting cyber attacks: (i) the changing trend of sensing data in a time segment is not preserved; 
(ii) the spatial patterns among sensors are captured partially;
(iii) the similarity between different data samples is not utilized completely. 

In order to overcome these limitations, we propose a new spatio-temporal graph (STG) to preserve and fuse spatio-temporal effects of WTNs simultaneously.
Moreover, a new pairwise kernel that utilizing the data similarity to augment the distance among different patterns is also proposed to improve the accuracy of cyber attack detection.

\section{Experimental Results}
We conduct experiments to answer the following research questions:
\begin{enumerate}[(1)]
    \item Does our proposed outlier detection framework (STOD) outperforms the existing methods?
    \item Is the spatio-temporal representation learning component  of STOD necessary for improving detection performance?
    \item Is the proposed pairwise kernel better than other traditional kernels for cyber attack detection?
    \item How much time will our method and other methods cost?
\end{enumerate}

\subsection{Data Description}
We used the secure water treatment system (SWAT) data set that is from Singapore University of Technology and Design for our study.
The SWAT has project built one water treatment system  and a sensor network to monitor and track the situations of the system.
Then, they construct one attack model to mimic the cyber attack of this kind of system in the real scenario.
The cyber attacks to and the sensory data of the system are collected to form the SWAT dataset. 
Table \ref{con:datades} show some important statistics of the SWAT dataset. 
Specifically, the SWAT dataset include a normal set (no cyber attacks)  and an attack set (with cyber attacks). The time period of the normal data is from 22 December 2015 to 28 December 2015. 
The time period of the attack data  is from 28 December 2015 to 01 January 2016, and 01 February 2016. 
There is no time period overlap between the normal data and the attack data on 28 January 2015.
It is difficult to identify more water treatment network datasets. In this study, we focus on validating our method using this dataset. 

\begin{table}[htbp]
\small
\centering
\setlength{\abovecaptionskip}{0.cm}
\caption{Statics of the SWAT data set}
\setlength{\tabcolsep}{1mm}{
\begin{tabular}{cccccc}  
\toprule
 Data Type  & Sensor Count & Total Items & Attack Items & Pos/Neg  \\  
\midrule       
  Normal  & 51 & 496800 & 0 & - \\
  Attack  & 51 & 449919 & 53900 & 7:1 \\
\bottomrule
\end{tabular}}
\label{con:datades}
 \vspace{-0.3cm}
\end{table}

\subsection{Evaluation Metrics}
We evaluate the performances of our method in terms of four metrics. 
Given a testset, a detection model will predict a set of binary labels (1: attack; 0: normal). 
Compared predicted labels with golden standard benchmark labels, we let $tp$, $tn$, $fp$, $fn$ be the sizes of the true positive, true negative, false positive, false negative sets, respectively. 

\begin{enumerate}[(1)]
\item \textbf{Accuracy:} is given by:
\begin{equation}
    Accuracy = \frac{tp+tn}{tp+tn+fp+fn}
\end{equation}

\item \textbf{Precision:} is given by:
\begin{equation}
    Precision = \frac{tp}{tp+fp}
\end{equation}

\item \textbf{F-measure:} is the harmonic mean of  precision and recall, which is given by:
\begin{equation}
    F-measure = \frac{2\times Precision \times Recall}{Precision+Recall}
\end{equation}

\item \textbf{AUC:} is the area under the ROC curve. It shows the capability of a model to distinguish between two classes.

\end{enumerate}

\begin{figure*}[!thb]
\setlength{\abovecaptionskip}{-8pt} 
	\centering
	\subfigure{\label{fig:accuracy}\includegraphics[width=0.24\linewidth]{{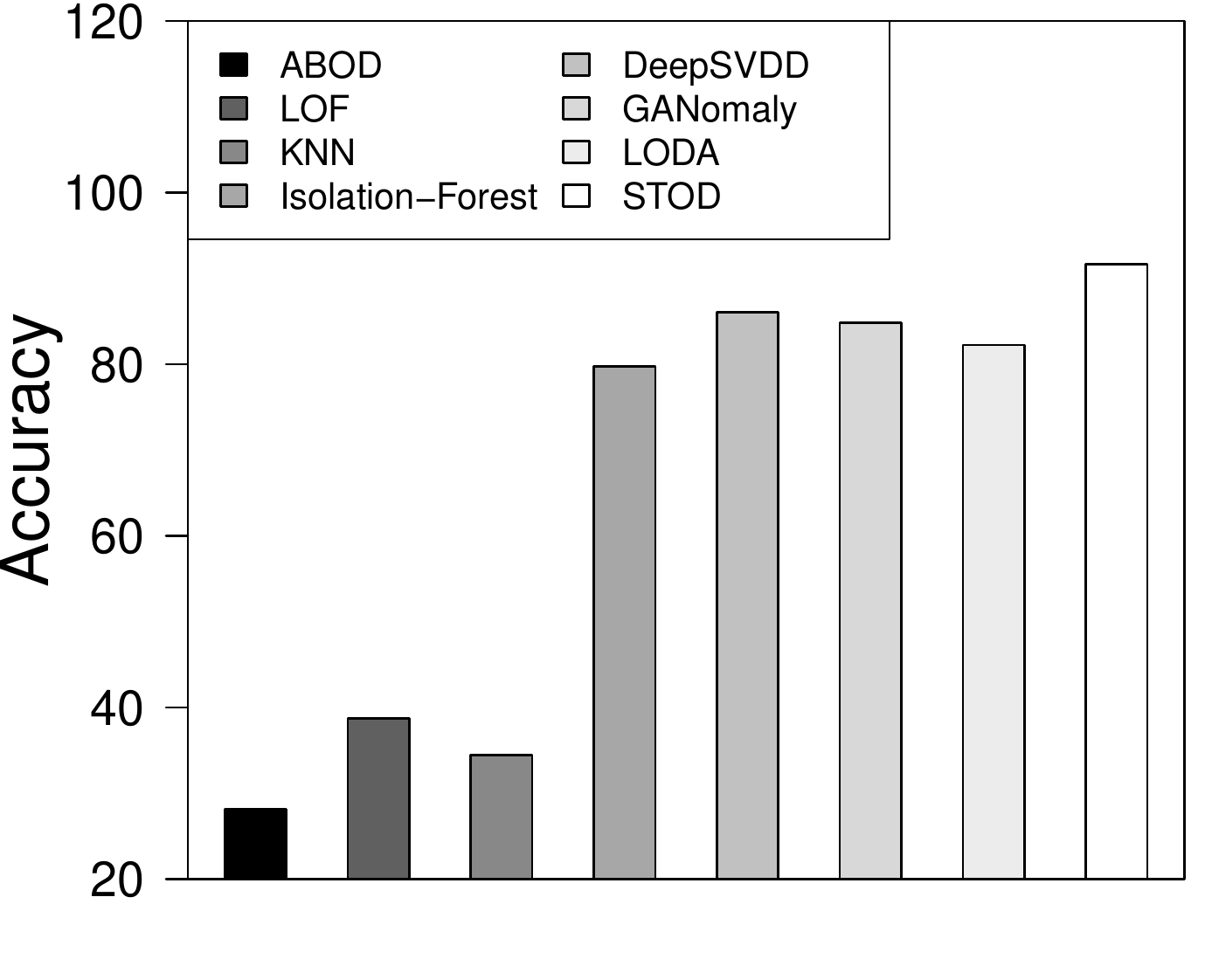}}}
	\subfigure{\label{fig:precision}\includegraphics[width=0.24\linewidth]{{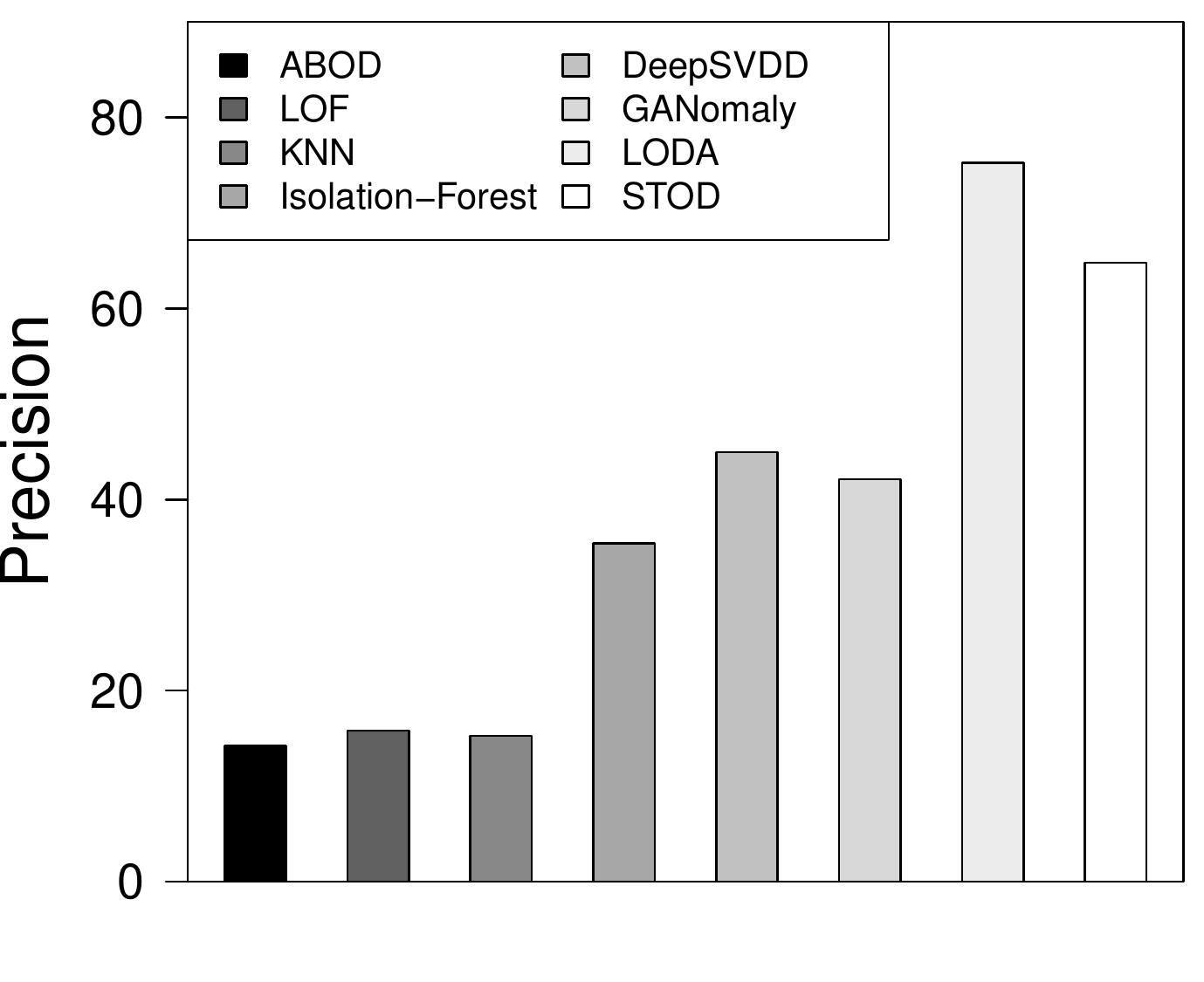}}}
	\subfigure{\label{fig:F1}\includegraphics[width=0.24\linewidth]{{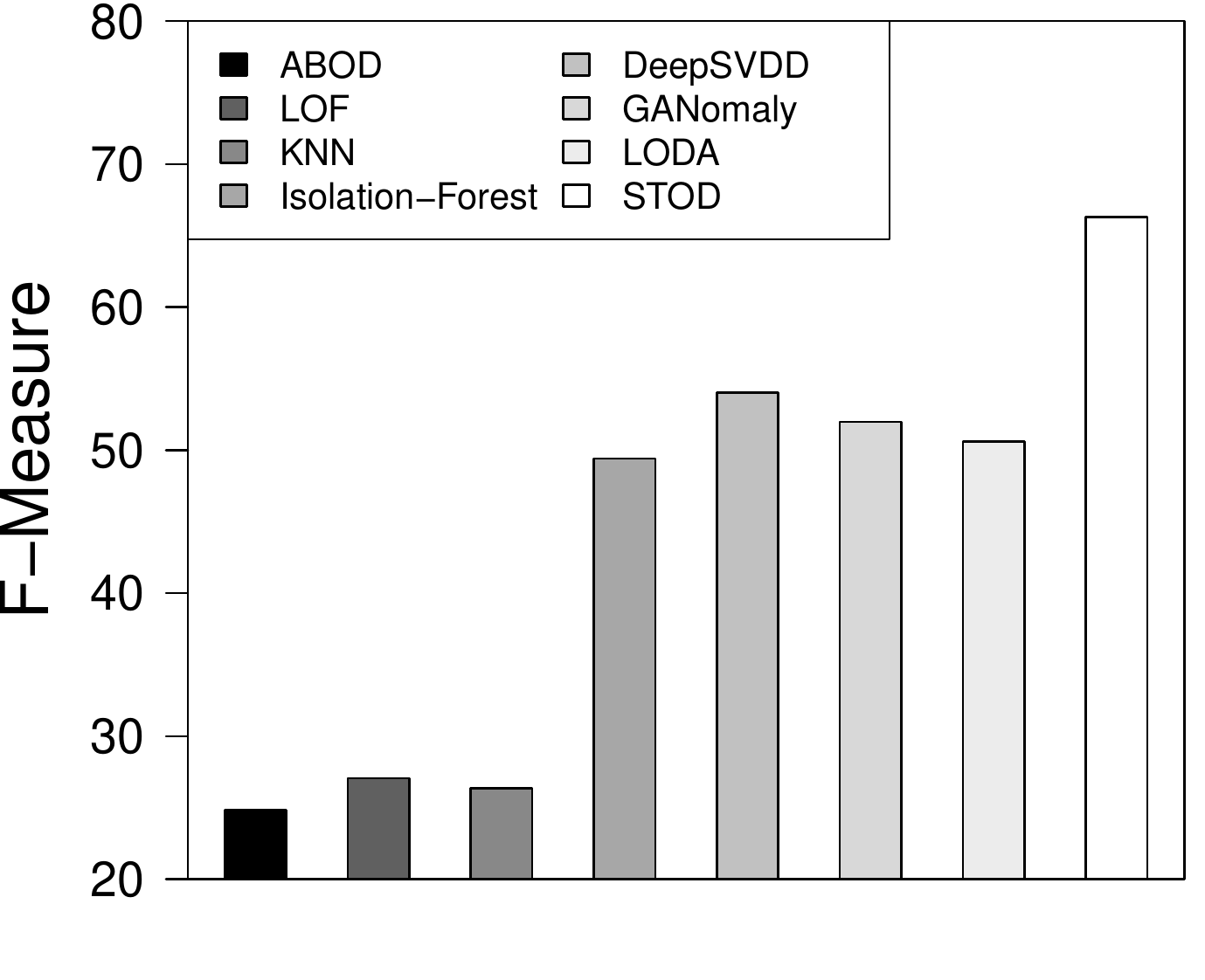}}}
	\subfigure{\label{fig:AUC}\includegraphics[width=0.24\linewidth]{{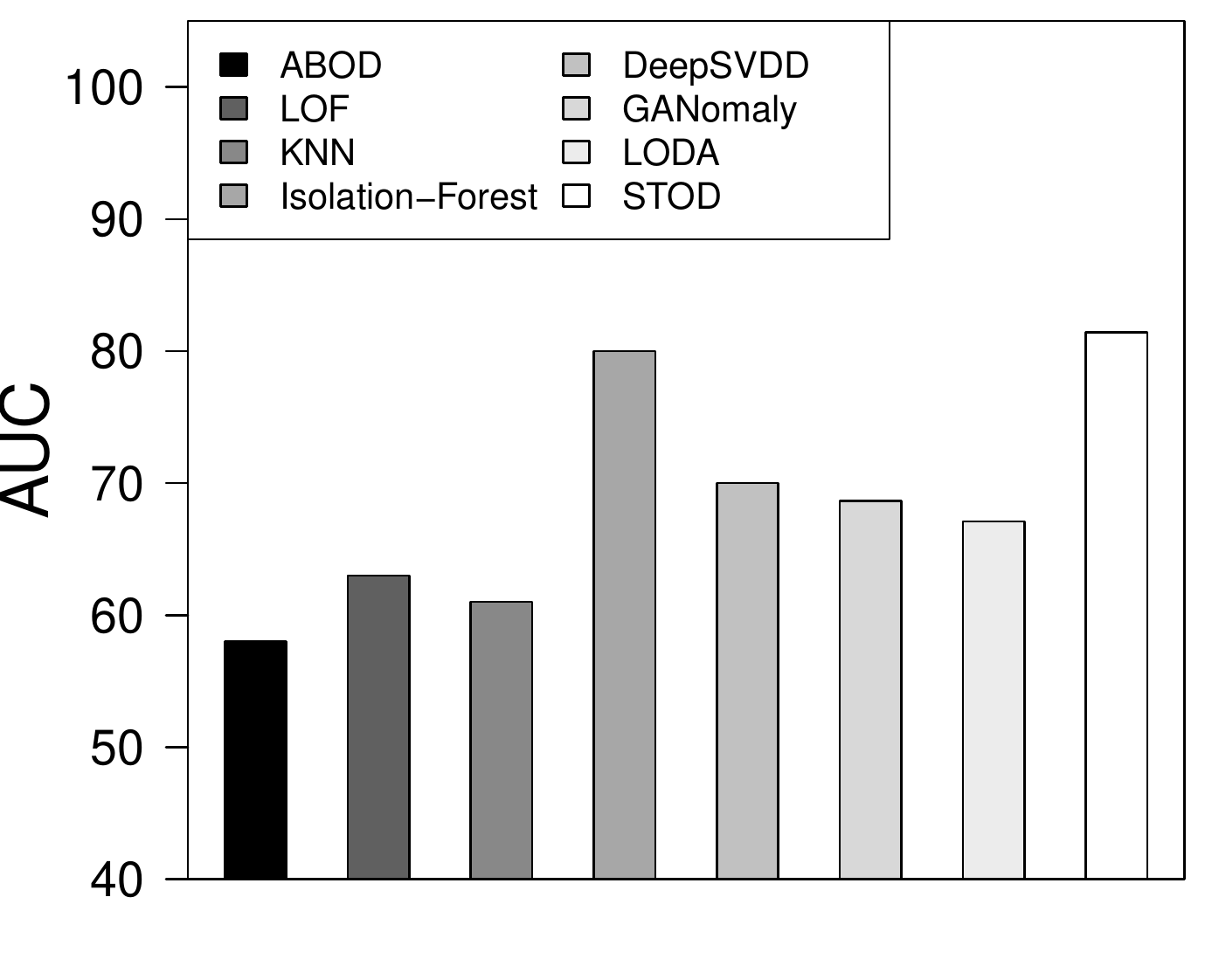}}}
	\caption{Comparison of different models in terms of Accuracy, Precision, F-measure and AUC .}
	\label{fig:overall}
\end{figure*}


\begin{figure*}[!thb]
\setlength{\abovecaptionskip}{-8pt} 
	\centering
	\subfigure{\label{fig:stod_accuracy}\includegraphics[width=0.24\linewidth]{{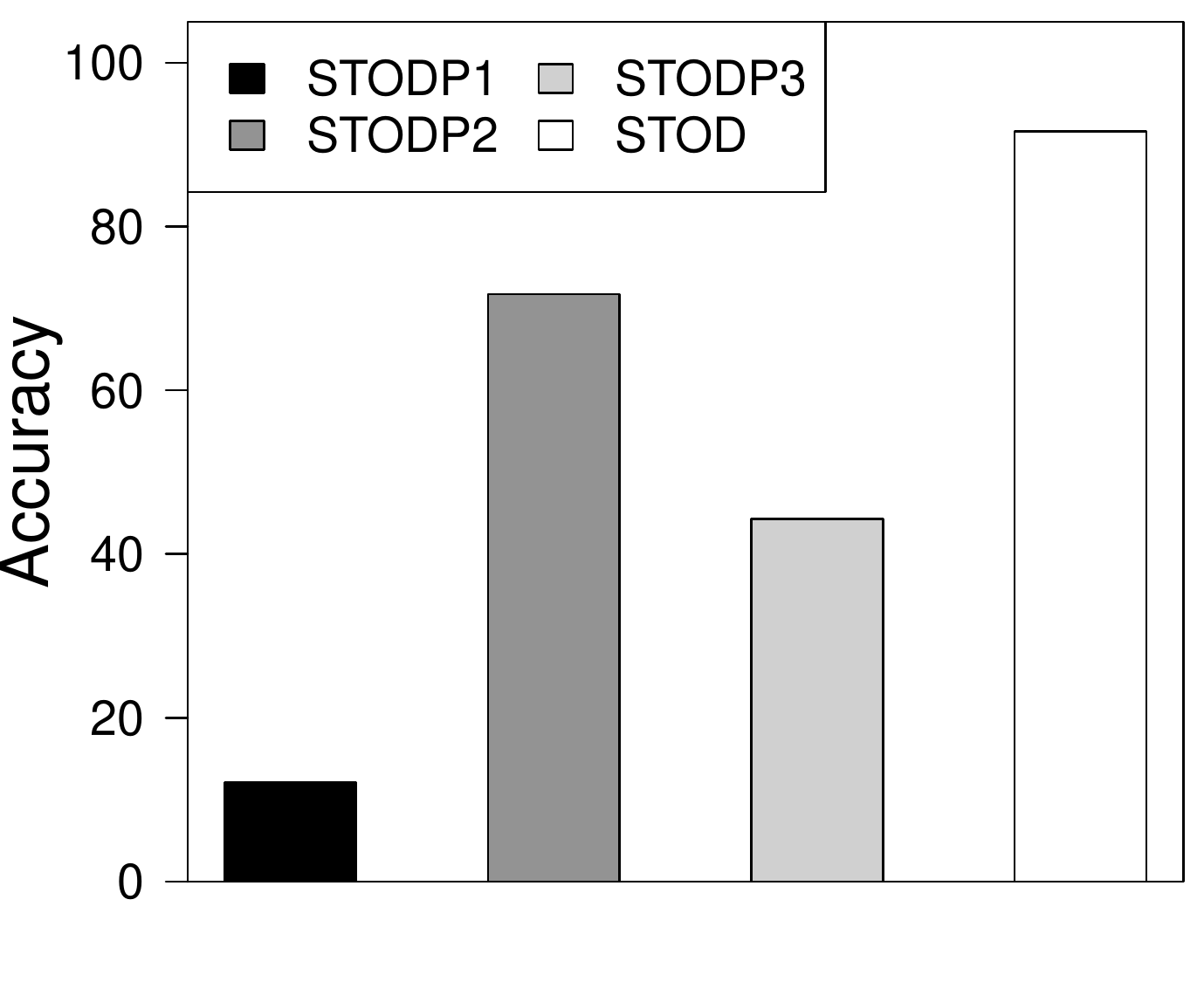}}}
	\subfigure{\label{fig:stod_precision}\includegraphics[width=0.24\linewidth]{{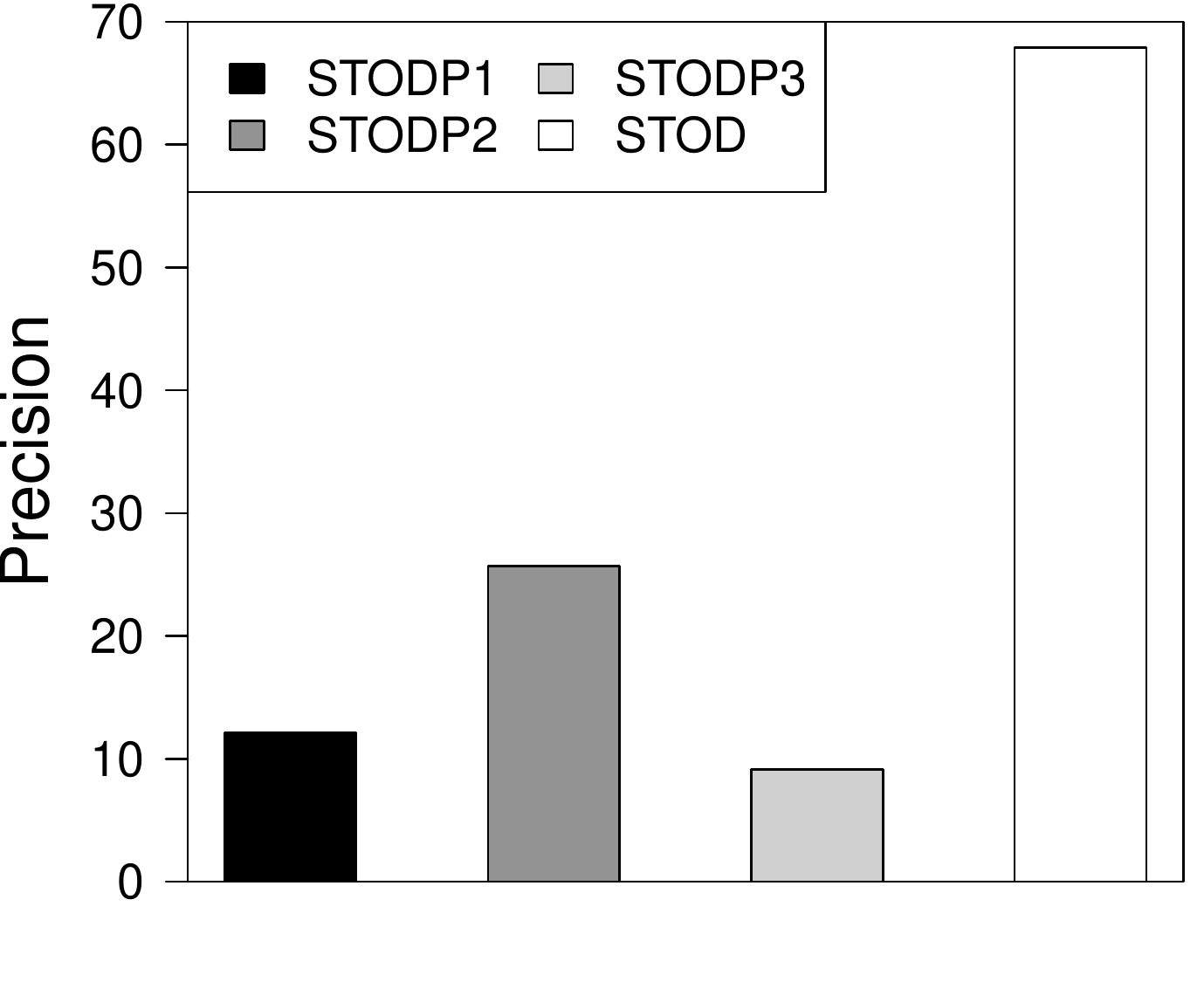}}}
	\subfigure{\label{fig:stod_F1}\includegraphics[width=0.24\linewidth]{{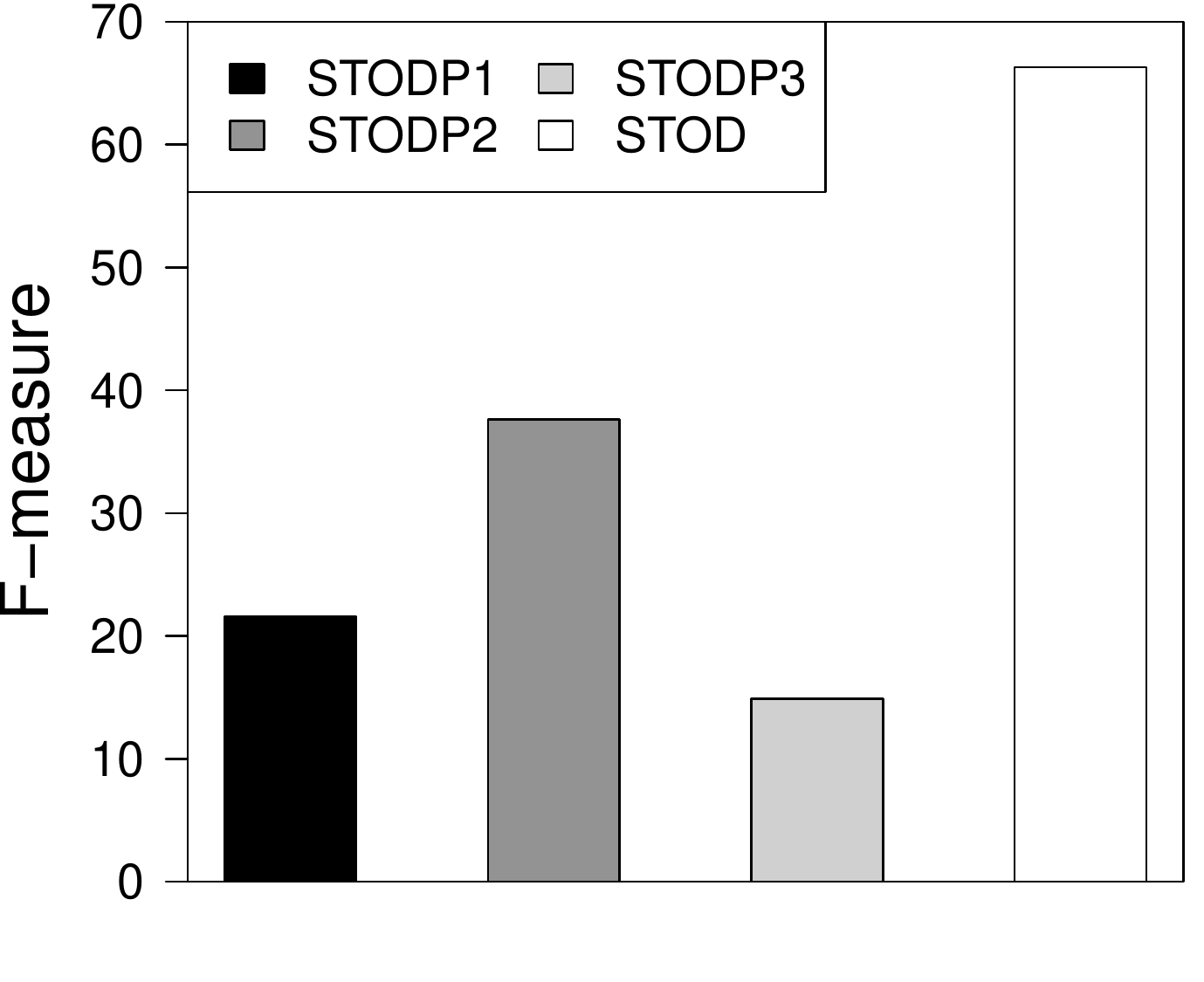}}}
	\subfigure{\label{fig:stod_AUC}\includegraphics[width=0.24\linewidth]{{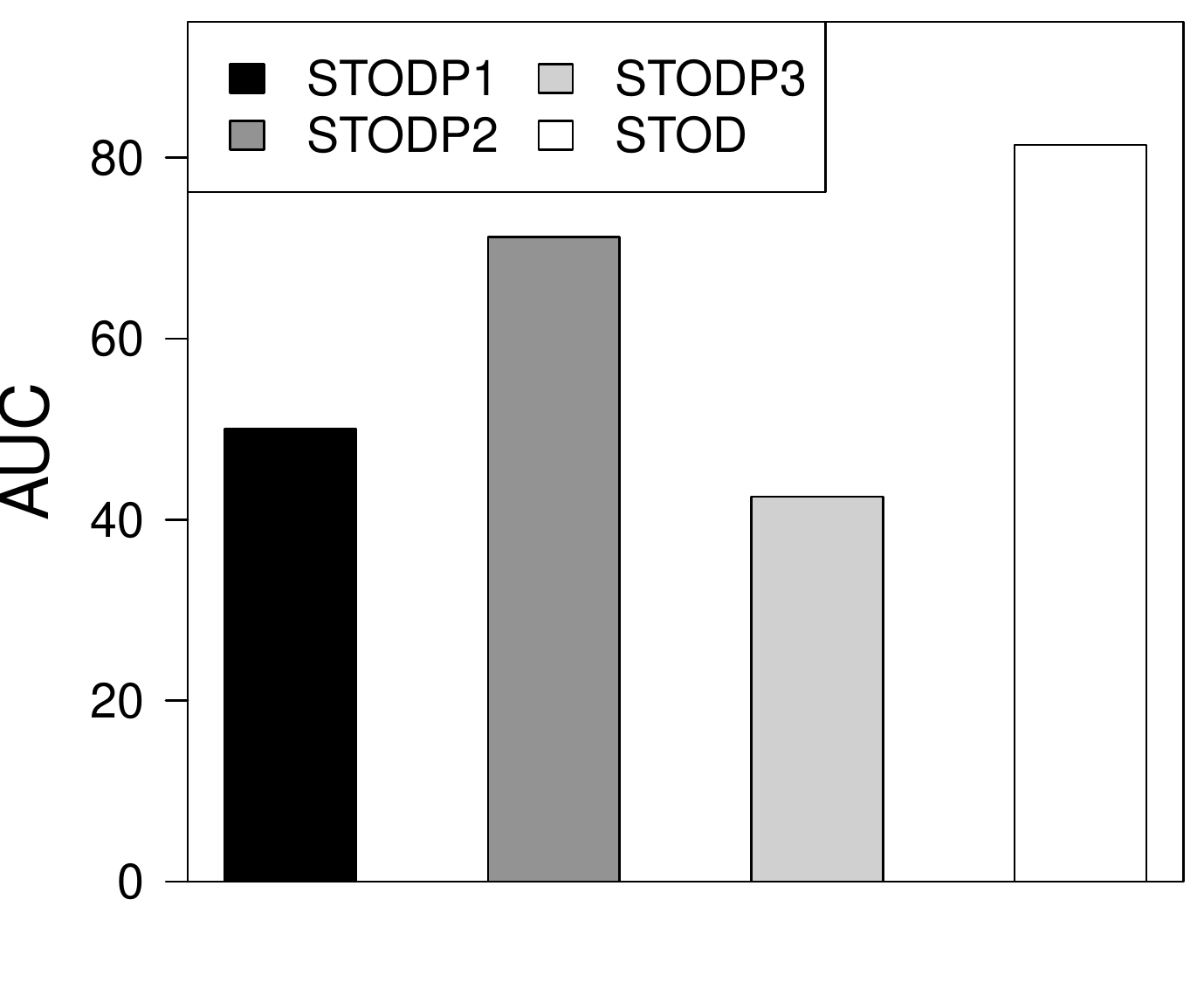}}}
	\caption{Comparison of different phases of representation learning module based on Accuracy, Precision, F-measure , and AUC.}
	\label{fig:stod}
 	\vspace{-0.5cm}
\end{figure*}

\subsection{Baseline Algorithms}
We compare the performances of our method (STOD) against the following ten baseline algorithms.

\begin{enumerate}[(1)]
    \item DeepSVDD \cite{ruff2018deep}:  expands the classic SVDD algorithm into a deep learning version. It utilizes a neural network to find the hyper-sphere of minimum volume that wraps the normal data. If a data sample falls inside of the hyper-sphere, DeepSVDD classifies the sample as  normal, and attack otherwise.
    In the experiments, we set the dimensions of the spatio-temporal embedding $\mathbf{z}_i$ to $28 \times 28$.
    
    \item GANomaly \cite{akcay2018ganomaly}: is based on the GAN framework.
    It develop a new version of generator by using the encoder-decoder-encoder structure. 
    The algorithm regards the difference between the embedding of the first encoder and the embedding of the second encoder as the anomaly score to distinguish normal and abnormal.
    In the experiments, we set the dimension of the spatio-temporal embedding vector $\mathbf{z}_i$ into $28 \times 28$.
    
    \item LODA \cite{pevny2016loda}: is an ensemble outlier detection model. It collects a series of weak anomaly detectors to produce a strong detector. In addition, the model fits real-time data flow and is resistant to missing values in the data set. In the experiments, we fed the learned representations into the LODA to detect.
    
    \item Isolation-Forest \cite{liu2008isolation}.
    The IsolationForest isolates observations by randomly selecting a feature and then randomly selecting a split value between the maximum and minimum values of the selected feature. In the experiments, we input spatio-temporal embedding vector $\mathbf{z}_i$ into Isolation-Forest, and set the number of estimators = 100, max sample numbers = 256.
    
    \item LOF \cite{breunig2000lof}. 
    The principle of LOF is to measure the local density of data samples. 
    If one data sample has low local density, the sample is an outlier. Otherwise, the sample is a normal sample. 
    In the experiments, we input the spatio-temporal embedding vector $\mathbf{z}_i$ into LOF and set the number of neighborhood = 20, the distance metric for finding neighborhoods is euclidean distance.
    
    \item KNN \cite{soucy2001simple}.
    KNN selects k nearest neighborhoods of one data sample based on a distance metric.
    KNN calculates the anomaly score of the data sample according to the anomaly situation of the $k$ neighborhoods. 
    In the experiments, we input spatio-temporal embedding vector $\mathbf{z}_i$ into KNN, and set the number of neighborhoods = 5, the adopted distance metric is euclidean distance.
    
    \item ABOD \cite{kriegel2008angle}.
    The ABOD method uses angle as a more robust measure to detect outliers. If many neighborhoods of one sample locate in the same direction to the sample, it is an outlier, otherwise, it is a normal sample.
    In the experiments, we input spatio-temporal embedding $\mathbf{z}_i$ into ABOD, set $k$ = 10. The angle metric is cosine value.  
    
    \item STODP1. We proposed to partition the sensing data into non-overlapped segments. The global mean pooling technique was then applied to fuse the segments of different sensors into an averaged feature vector. We fed the fused feature vector  into OC-SVM for outlier detection. 
    The kernel of OC-SVM is defined in Equation \ref{con:newk}.
    
    \item STODP2. We applied the global mean pooling 
    to the temporal embedding vectors generated by Section 3.A to obtain a global feature vector of WTN, which was fed into OC-SVM for outlier detection.
    In addition, the kernel of OC-SVM is our proposed kernel function defined in Equation \ref{con:newk}.
    
    \item STODP3. In order to study the effect of Seq2Seq, we remove the Seq2Seq module of our framework pipeline.
    The temporal segments of different sensors are organized as graph set.
    The graph set is input into graph embedding module to obtain the final embedding.
    Finally, the embedding is input into OC-SVM to do outlier detection.
    The kernel of the OC-SVM is  defined in Equation \ref{con:newk}.

\end{enumerate}

In the experiments, the spatio-temporal representation learning phase of our framework is used to preserve the spatio-temporal patterns and data characteristics into feature learning. 
The one-class outlier detection phase of our framework is used to detect the cyber attack status of the water treatment system based on the spatio-temporal representation.
We only use normal data to train our model. 
After the training phase, our model has the capability to detect the status of the testing data set that contains both normal and attack data.
All the evaluations are performed on a x64 machine with Intel i9-9920X 3.50GHz CPU and 128GB RAM. The operating system is Ubuntu 18.04.

\subsection{Overall Performances}

We compare our method with the baseline models in terms of accuracy, precision, f-measure and AUC. 
Figure \ref{fig:overall} shows the average performances of our mtehod (STOD) is the best in terms of accuracy, f-measure and AUC; our method ranks second in terms of precision, compared with other baseline models. 
A potential interpretation of such observation is that the STOD captures the temporal effects (\textbf{delayed} , \textbf{continued}) and spatial effect (\textbf{cascading}) of cyber attacks by spatio-temporal representation learning part of STOD in a balanced way. 
With STOD captures more intrinsic features of cyber attacks, the model not only finds more attack samples but also makes fewer mistakes on normal samples.
Thus, the distinguishing ability of STOD is improved greatly.
But on a single evaluation metric, STOD maybe poorer than other baselines.
Overall, STOD outperforms with respect to Accuracy, F-measure and ACU compared with baseline models, which signifies our detection framework owns the best attack detection ability.

\begin{figure*}[!thb]
\setlength{\abovecaptionskip}{-8pt} 
	\centering
	\subfigure{\label{fig:kernel_accuracy}\includegraphics[width=0.24\linewidth]{{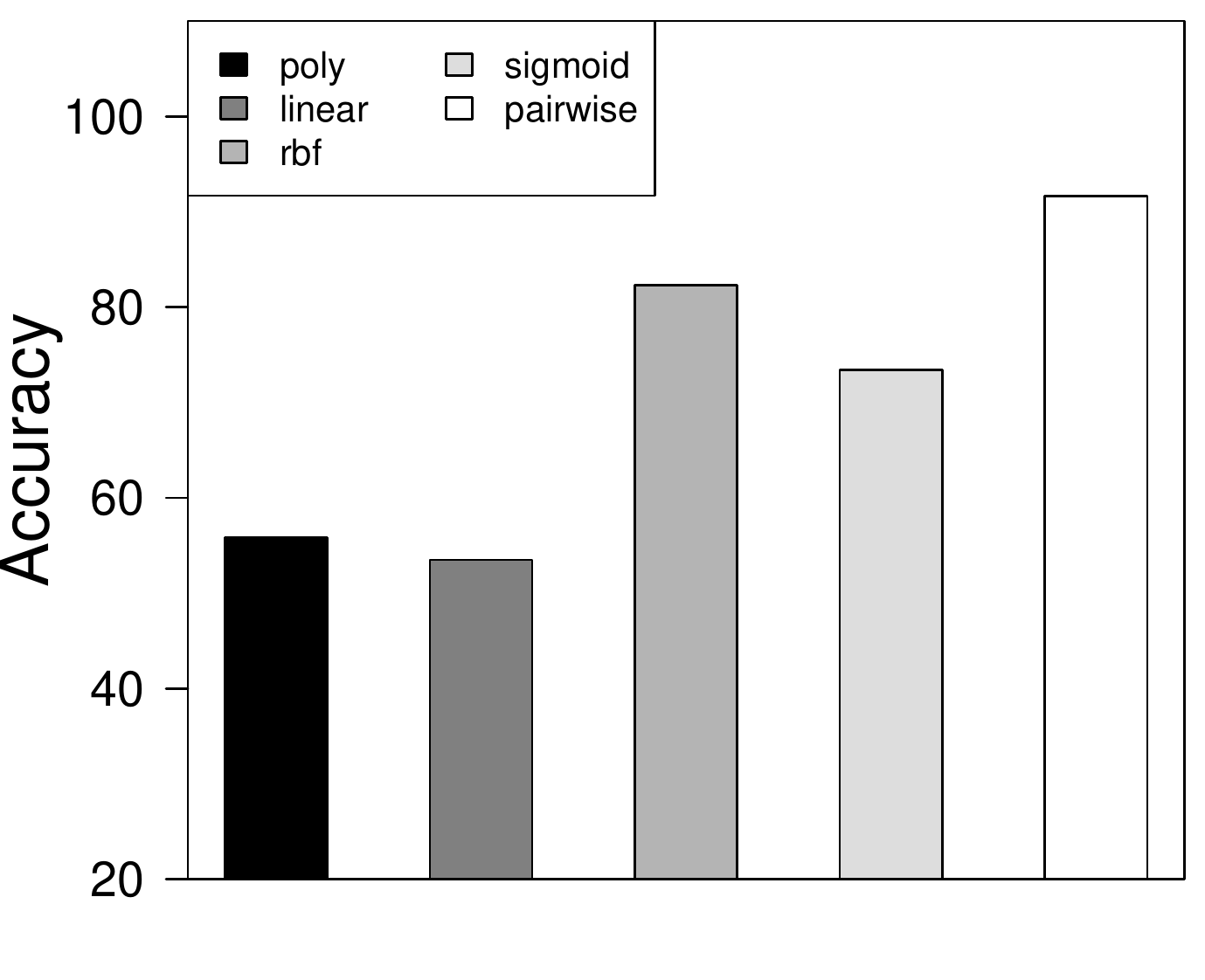}}}
	\subfigure{\label{fig:kernel_precision}\includegraphics[width=0.24\linewidth]{{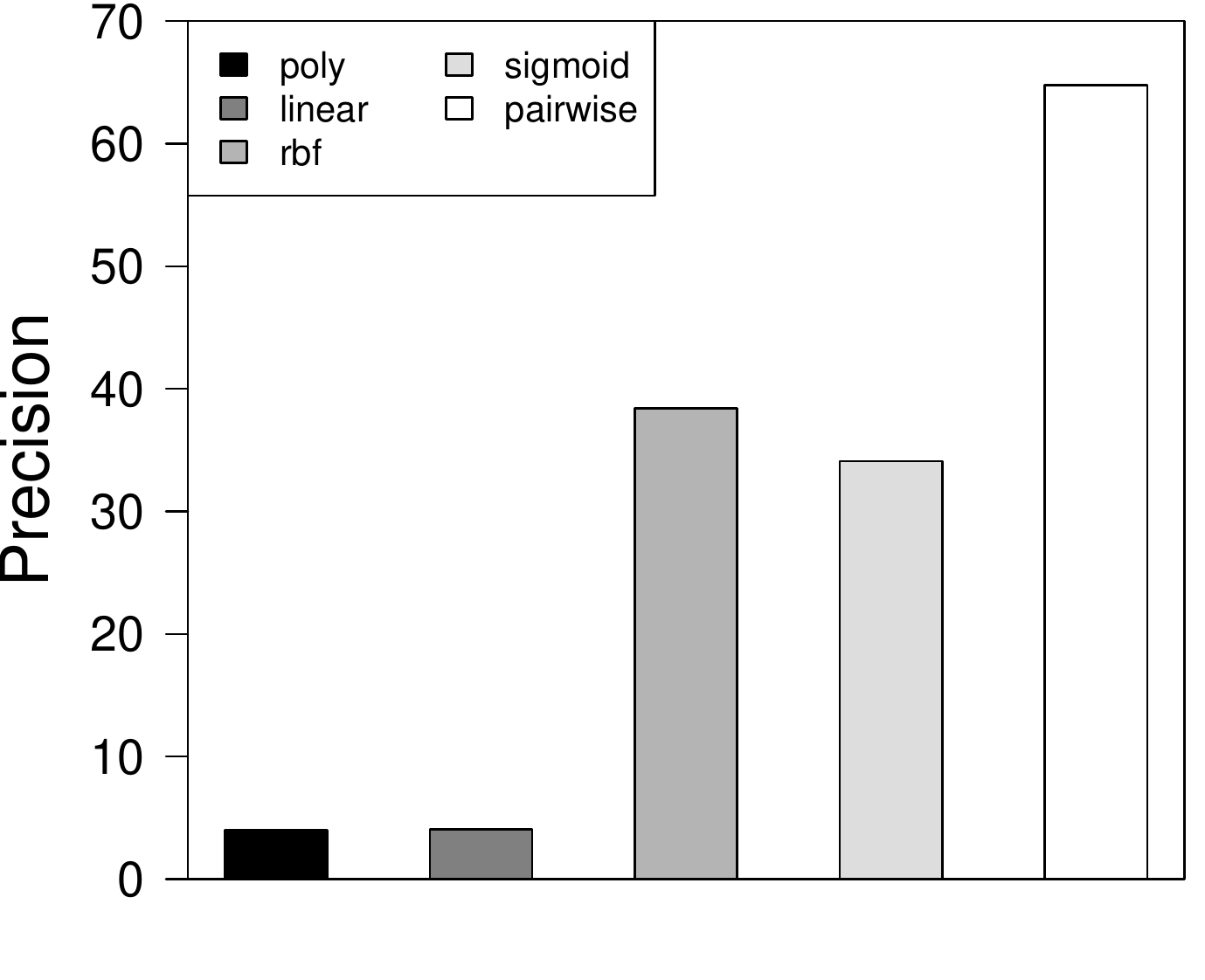}}}
	\subfigure{\label{fig:kernel_F1}\includegraphics[width=0.24\linewidth]{{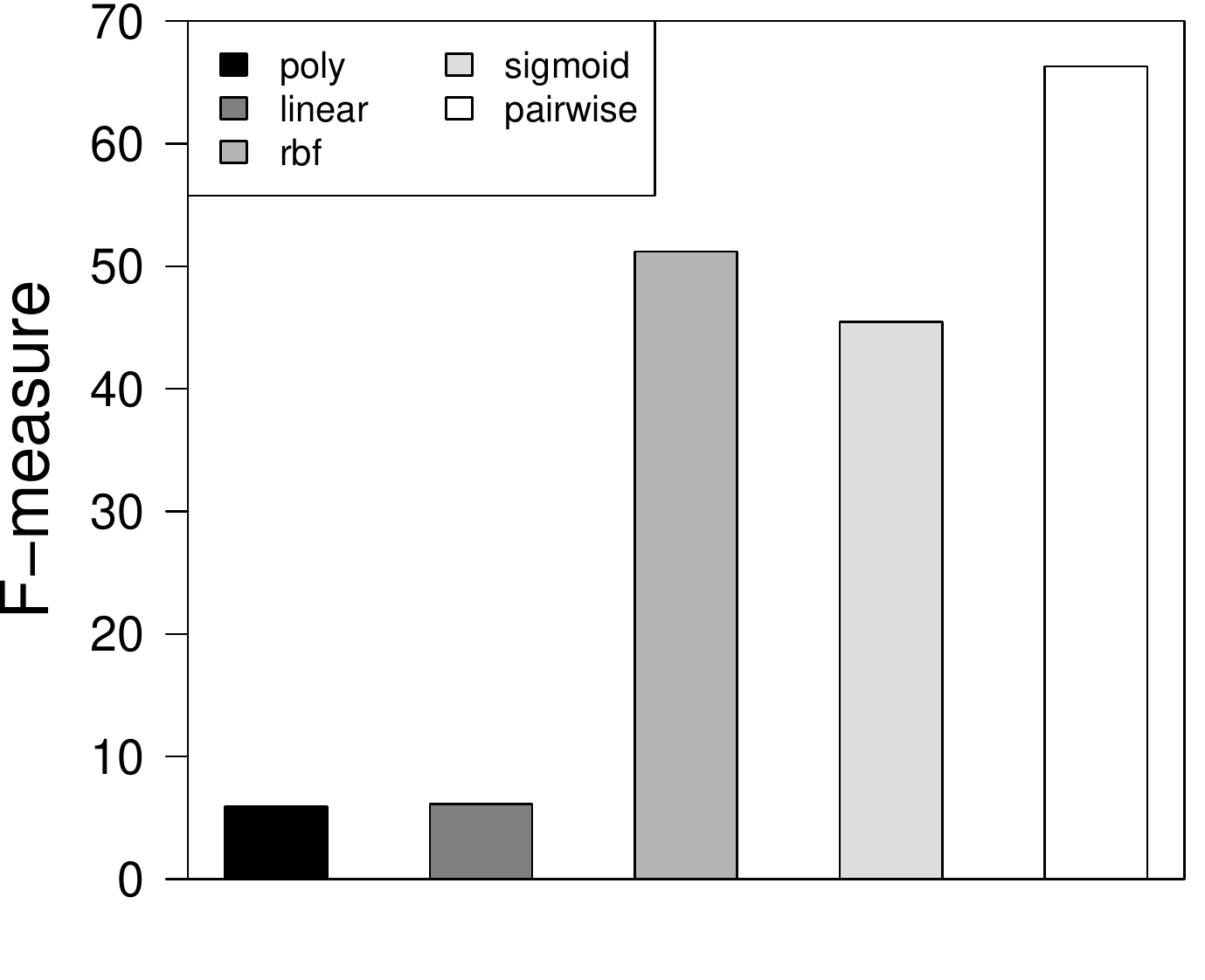}}}
	\subfigure{\label{fig:kernel_AUC}\includegraphics[width=0.24\linewidth]{{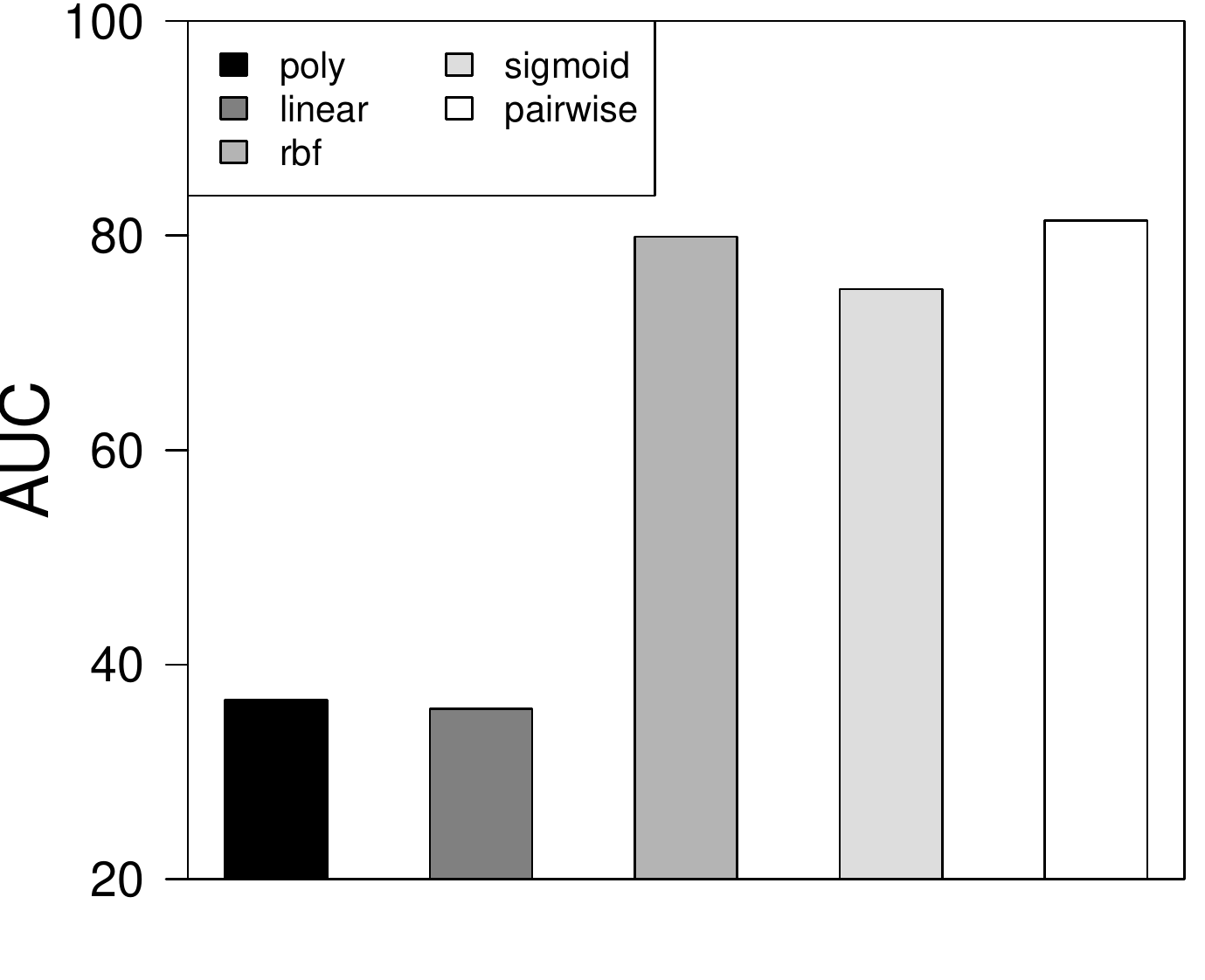}}}
	\caption{Comparison of different kernels with respect to Accuracy, Precision, F-measure , and AUC.}
	\label{fig:kernel_func_exp}
 	\vspace{-0.5cm}
\end{figure*}

Another observation is that the performances of LOF, ABOD, and KNN are much worse than other models. 
The possible reason is that these models exploit distance or angle-based assessment strategies.
These geometrical measurements are vulnerable after projecting data into high dimensional space due to the ``curse of dimensionality''.
Thus, these models can not achieve excellent performances.

\subsection{Study of Representation Learning}

The representation learning phase of our framework include:
(1) partitioning sensor data streams into segments;
(2) modeling the temporal dependencies with seq2seq;
(3) modeling the spatial dependencies with graph embedding.
What role does each of the three steps play in our framework?
We will iteratively remove each of the three steps to obtain three different variants, namely STODP1, STODP2, STODP3. 
We then test compare the three variants with our original framework to examine the importance of the removed step for improving detection performances

Figure \ref{fig:stod} shows the experimental results of STOD, STODP1, STODP2, and STODP3, which clearly show that STOD outperforms STODP1, STODP2, and STODP3 in terms of accuracy, precision, f-measure, and AUC with a large margin.
A reasonable explanation of this phenomenon is that attack patterns are spatially and temporally structured, and, thus, when more buried spatio-temporal patterns are modeled, the method becomes more discriminant. 
The results validate the three steps (segmentation, temporal, spatial) of the representation learning phase is critical for attack pattern characterization. 

\subsection{Study of Pairwise Kernel Function }

The kernel function is vital for the SVM based algorithm family. An effective kernel function can map challenging data samples into a high-dimensional space, and make these data samples more separable in the task of detection. 
We design experiments to validate the improved performances of our pairwise kernel function by comparing our pairwise kernel function with  other baseline kernel functions.
Specifically, the baseline kernels are as follows:
\begin{enumerate}[(1)]
    \item \textbf{linear}. This kernel is a linear function. There are limited number of parameters in the linear kernel, so the calculation process is quick. The dimension of the new feature space is similar to the original space.
    \item \textbf{poly}. This kernel is a polynomial function.
    The parameters of the kernel are more than the linear kernel.
    It maps data samples into high dimensional space.
    \item \textbf{rbf}. This kernel is a Gaussian function that is a non-linear function. It exhibits excellent performance in many common situations.
\item \textbf{sigmoid}. This kernel is a sigmoid function.
When SVM utilizing this function to model data samples, the effect is similar to using a multi-layer perceptron. 
    
\end{enumerate}

Figure \ref{fig:kernel_func_exp} shows a comparison between our kernel and other baseline kernels with respect to all evaluation metrics. 
We observed that our kernel shows significant improvement, compared with other baseline kernels, in terms of Accuracy, Precision, F-measure, and AUC, 
This indicates that our kernel can effectively augment the attack patterns in original data, and maximize the difference between normal and attack patterns, by mapping original data samples into high dimensional feature space. 
This experiment validates the superiority of our pairwise kernel function.

\subsection{Study of Time Costs}
We aim to study the time costs of  training and testing in different models. 
Specifically, we divided the dataset into six non-overlap folds. 
We then used cross-validation to evaluate the time costs of different models.

\begin{figure}[htbp]
    \setlength{\abovecaptionskip}{-0.1cm} 
    \centering
    \includegraphics[width=0.4\textwidth]{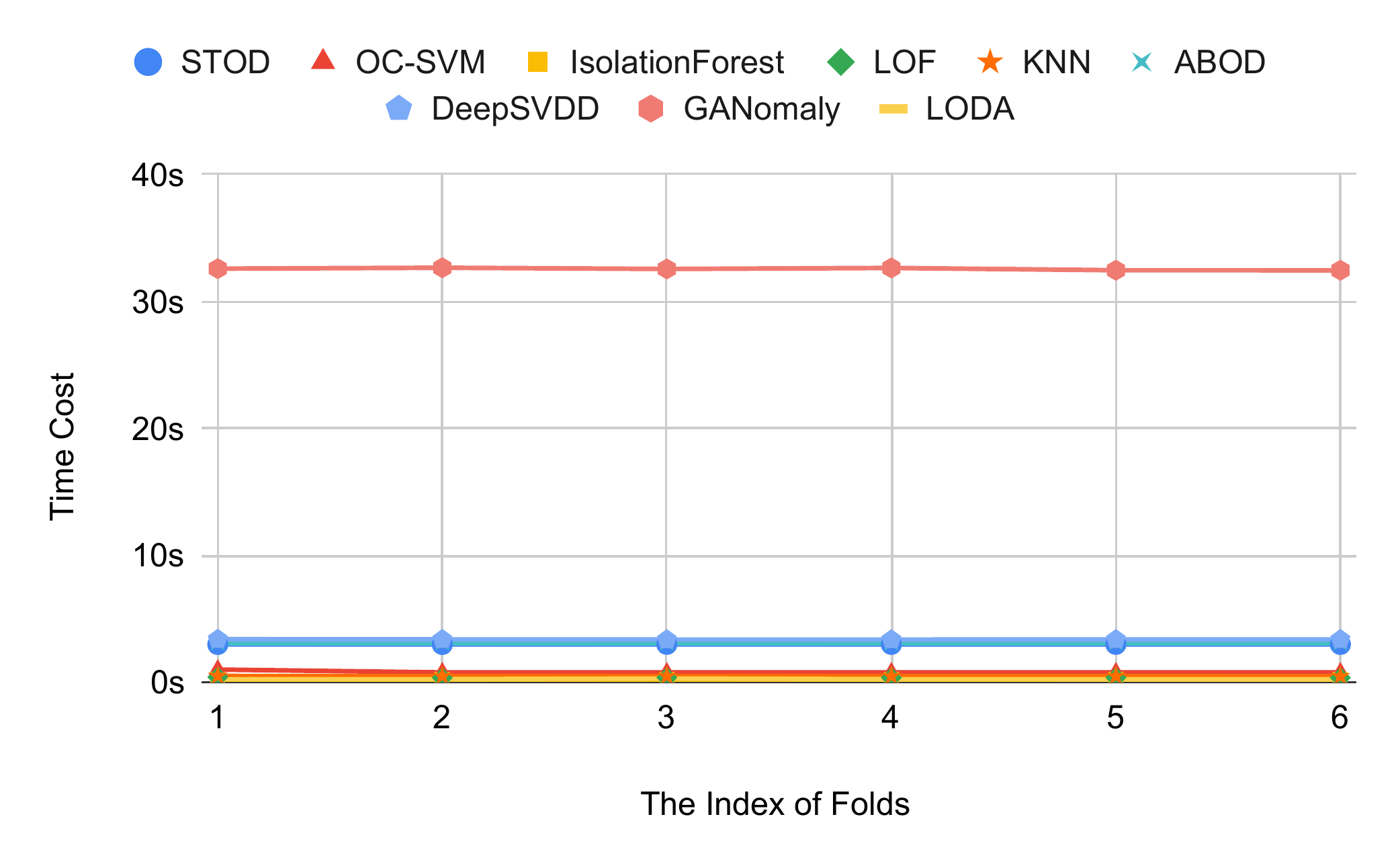}
    \caption{Comparison of different models based on training time cost}
    \label{fig:fit_time}
    \vspace{-0.4cm}
\end{figure}

Figure \ref{fig:fit_time} shows the comparison of training time costs among  different models. 
We find that the training time costs of each model is relatively stable. 
An obvious observation is GANomaly has the largest training time cost than other models. This is because the encoder-decoder-encoder network architecture design is time-consuming.
In addition, the training time of STOD is slightly longer than OC-SVM. 
This can be explained by the fact that the similarity calculation of pairwise kernel function increases time costs, since we need to calculate the similarities between two representation vectors of each training data sample.

\begin{figure}[htbp]
    \setlength{\abovecaptionskip}{-0.1cm} 
    \centering
    \includegraphics[width=0.4\textwidth]{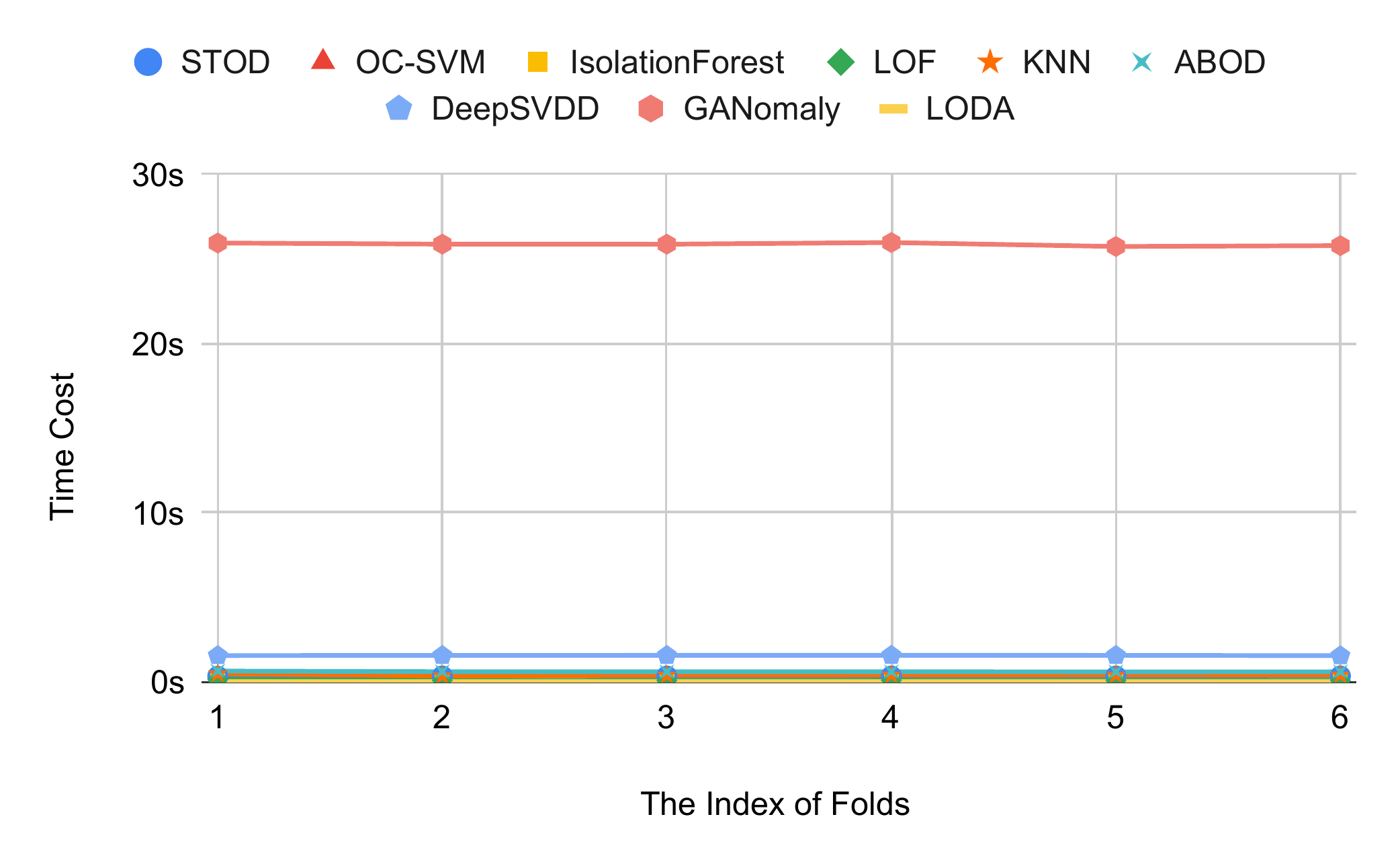}
    \caption{Comparison of different models based on testing time cost.}
    \label{fig:score_time}
     \vspace{-0.3cm}
\end{figure}
Figure \ref{fig:score_time} shows the comparisons of testing time costs among different models. 
The testing time costs of each model are relatively stable as well. And  many of them can complete the testing task within one second,  except GANomaly. 
We find the testing time of our method is shorter than the training time by comparing Figure \ref{fig:fit_time} and Figure \ref{fig:score_time}.
This can be explained by a strategy of our method:  once the model training is completed, our method stores the kernel mapping parameters in order to save time of computation. 
In addition, GANomaly still shows the largest testing time cost.
The reason is that the testing phase of GANomaly needs to calculate two representation vectors of each testing data samples, and, thus, GANomaly doesn't use less time, compared with that of the training phase.

\subsection{Case Study: Visualization for Spatio-temporal Embedding}
\begin{figure}[htbp]
    \setlength{\abovecaptionskip}{-0.1cm} 
    \centering
    \includegraphics[width=0.3\textwidth]{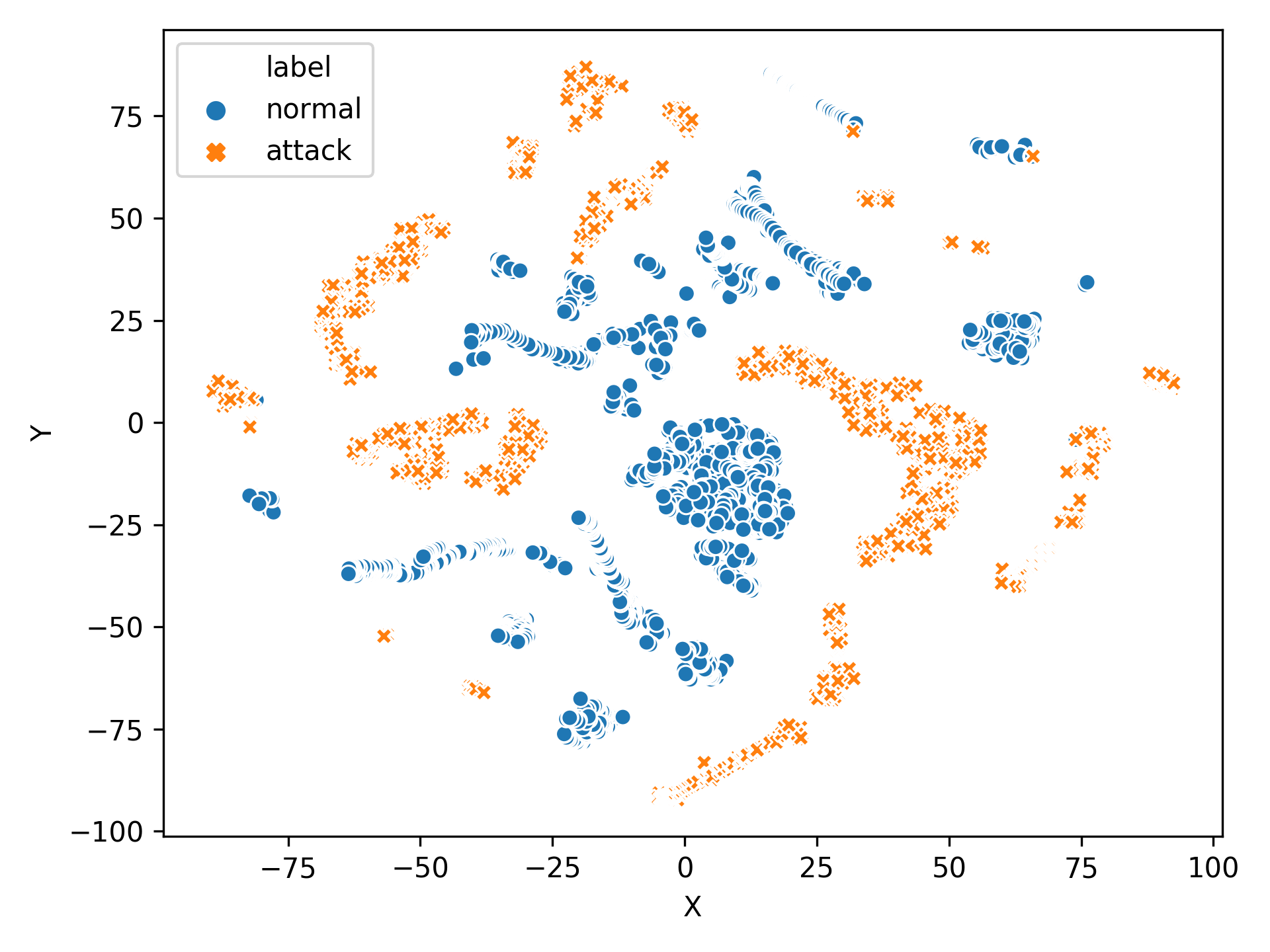}
    \caption{visualization result for spatio-temporal embedding}
    \label{fig:emb_st_v}
     \vspace{-0.3cm}
\end{figure}

The spatio-temporal representation learning phase is an important step in our structured detection framework. 
An effective representation learning method should be able to preserve the patterns of normal or attack behaviors and maximize the distances between normal and attack in the detection task.
We visualize the spatio-temporal embedding on a 2-dimensional space, in order to validate the discriminant capabilities of our learned representations. 
Specifically, we first select 3000 normal and 3000 attack spatio-temporal embedding respectively.
We then utilize the T-SNE manifold method to visualize the embedding.
Figure \ref{fig:emb_st_v} shows the visualization results of normal and attack data samples.
We find that our representation learning result is discriminant in a transformed 2-dimensional space.
As can be seen, the learned normal and attack representation vectors are clustered together to form dense areas. The observation shows that non-linear models are more appropriate for distinguishing normal and attack behaviors than linear methods.

\section{Related Work}

\textbf{Representation Learning}. Representation learning is to learn a low-dimensional vector to represent the given data of an object. 
Representation learning approaches are three-fold: (1) probabilistic graphical models; (2) manifold learning approaches; (3) auto-encoder and its variants;
The main idea of the probabilistic graphical model is to learn an uncertain knowledge representation by a Bayesian network~\cite{friedman2004inferring,johnson2016composing}.
The key challenge of such methods is to find the topology relationship among nodes in the probabilistic graphical model.
The manifold learning methods utilize the non-parametric approach to find manifold and  embedding vectors in low dimensional space based on neighborhood information~\cite{zhu2018image,wang2018flexible}. 
However, manifold learning is usually time-costly. 
The discriminative ability of such methods is very high in many applications. 
Recently, deep learning models are introduced to conduct representation learning.
The auto-encoder model is a classical neural network framework, which embeds the non-linear relationship in feature space via minimizing the reconstruction loss between original and reconstructed data~\cite{wang2016auto,suk2015latent,calvo2019selectional}. 
When representation learning meets spatial data, autoencoders can inetgrate with spatio-temporal statistical correlations to   learn more effective embedding vectors.  ~\cite{cedolin2010spatiotemporal,ma2019ts,pan2019urban}. 
For instance, Singh et al, use the auto-encoder framework to learn the spatio-temporal representation of traffic videos to help detect the road accidents~\cite{singh2018deep}. 
Wang et al. utilize spatio-temporal representation learning to learn the intrinsic feature of GPS trajectory data to help analyze  driving behavior~\cite{wang2019spatiotemporal}.

\textbf{Deep Outlier Detection}.  Outlier detection is a classical problem with important applications, such as,  fraud detection and cyber attack detection.
Recently, deep learning has been introduced into outlier detection.
According to the availability of outlier labels, deep anomaly detection can be classified into three categories:
(1) supervised deep outlier detection;
(2) semi-supervised deep outlier detection;
(3) unsupervised deep outlier detection.
First, supervised deep outlier detection models usually train a deep classification model to distinguish whether  a data sample is normal or not~\cite{yamanaka2019autoencoding,kawachi2019two}.
These models are not widely available in reality,  because it is difficult to obtain data labels of outliers. 
Meanwhile, data imbalance is a serious issue that degrades the performances of supervised models. 
Second, semi-supervised outlier detection methods usually train a deep auto-encoder model to learn the latent embedding of normal data \cite{pmlr-v80-ruff18a,chalapathy2018anomaly,zhao2016classification}, then the learned embedding vectors are used to accomplish outlier detection task.
In deep semi-supervised outlier detection, one-class classification is an important research direction.
For instance, Liu et. al, proposed to detect the anomaly data on uncertain data by SVDD algorithm \cite{liu2013svdd}. Many experiments have shown the adaptability of one class SVM. 
Third, unsupervised outlier detection models do not need any label information, they detect outliers depends on the intrinsic rules (e.g., scores, distance, similarity) of data~\cite{liu2019generative,wang2019effective,lu2017unsupervised}.
Such methods are appropriate for scenarios that are hard to collect label information.

\textbf{Cyber Attack Detection in Water Treatment Network}. Water purification plants are critical infrastructures in our local communities. 
Such infrastructures are usually vulnerable to cyber attacks. 
Early detection of cyber attacks in water treatment networks is significant for defending our infrastructure safety and public health.  
There are many existing studies about outlier detection in water treatment networks~\cite{feng2017multi,romano2010real,lin2018tabor,ramotsoela2019attack,adepu2016using}. 
For instance, Adepu et al. studied the impact of cyber attacks on water distribution systems~\cite{adepu2019investigation}. 
Goh et al. designed an unsupervised learning approach that regards Recurrent Neural Networks as a temporal predictor to detect attacks \cite{goh2017anomaly}.
Inoue et al. compared the performances of Deep Neural Network and OC-SVM on outlier detection in water treatment networks~\cite{inoue2017anomaly}. 
Raciti et al. developed a real-time outlier detection system by  clustering algorithms and deployed the system into a water treatment network~\cite{raciti2012anomaly}.
However, there is limited studies that integrate deep graph representation learning, spatiotemporal patterns, and one-class detection to more effectively address cyber attack problems.

\section{Conclusion Remarks}
We studied the problem of cyber attack detection in water treatment networks. 
To this end, we proposed a structured detection framework to integrate spatial-temporal patterns, deep representation learning, and one-class detection. 
Specifically, we first partitioned the sensing data of WTNs into a sequence of fixed-size time segments. 
We then built a deep spaiotemporal representation learning approach to preserve the spatio-temporal patterns of attacks and normal behaviors.
The representation learning approach includes two modules:
(1) a temporal embedding module, which preserves the temporal dependencies within a time segment.
Then, we constructed the spatiotemporal graphs by mapping the temporal embedding to the WTN as node attributes.
(2) a spatial embedding module, which learns the fused spatio-temporal embedding from the spaiotemporal graphs.
In addition, we developed an integrated one-class detection method with an improved pairwise kernel.
The new kernel is capable of augmenting the difference between normal and attack patterns via the pairwise similarity among deep embedding vectors of system behaviors. 
Finally, we conducted extensive experiments to illustrate the effectiveness of our method: STOD achieves an accuracy of $91.65\%$, with average improvement ratios of $82.78\%$  and $22.96\%$ with respect to F1 and AUC, compared with the baseline methods.

\normalem
\bibliographystyle{IEEEtran}    
\bibliography{ref}

\end{document}